\documentclass[twocolumn,showpacs,preprintnumbers,amsmath,amssymb,superscriptaddress]{revtex4-1}
\usepackage{graphicx}
\usepackage{dcolumn}
\usepackage{bm}
\usepackage[english]{babel}

\begin{document}

\title{Generalized vortex-model for the inverse cascade of two-dimensional turbulence}
\author{J.~Friedrich}
\email{Electronic mail: jaf@tp1.rub.de}
\affiliation{Institute for Theoretical Physics I, Ruhr-University Bochum, Universit\"{a}tsstr. 150,
D-44801 Bochum, Germany}
\author{R.~Friedrich}
\affiliation{Institute for Theoretical Physics, University of M\"unster,
Wilhelm-Klemm-Str. 9, D-48149 M\"unster, Germany}
\date{\today}
\begin{abstract}
We generalize Kirchhoff's point vortex model of two-dimensional fluid motion
to a rotor model which exhibits an inverse cascade
by the  formation of rotor clusters. A rotor is composed of two vortices with
like-signed circulations glued together by an overdamped spring. 
The model is motivated by a treatment of the vorticity equation 
representing the vorticity field 
as a superposition of vortices with elliptic 
Gaussian shapes of variable widths, 
augmented by a suitable forcing mechanism. The rotor model opens up the way
to discuss the energy transport in the inverse cascade on the basis of
dynamical systems theory.
\end{abstract}
\maketitle
\section{Introduction}
The theoretical treatment of the longstanding problem of turbulent flows
\cite{Monin,Frisch,Tsinober,FalkSreen} 
has to relate dynamical systems theory with 
non-equilibrium statistical physics \cite{CFG}.
The central notion of physical turbulence
theory is the concept of the
energy cascade, highlighting the fact that turbulent flows are essentially
transport processes of quantities like energy or enstrophy 
in scale. Although well-established
theories due to Richardson, Kolmogorov, Onsager, Heisenberg
and others (for reviews we refer the reader 
to \cite{Monin,Frisch,Tsinober})
can capture gross features of the cascade process in a 
phenomenological way, the dynamical aspects are by far less understood, 
and usually are investigated by direct numerical simulations of the Navier-Stokes equations.

An exception, in some sense, are inviscid fluid flows in two
dimensions. Based on the work of Helmholtz \cite{Helmholtz}, it was 
Kirchhoff \cite{Kirchhoff} who pointed out that the partial
differential equation can be reduced to a Hamiltonian system for 
the locations of point vortices, provided one considers initial conditions 
where the vorticity
is a superposition of delta-distributions (we refer the reader to
the works of Aref \cite{Aref1,Aref2,Aref3} as well as the monographs \cite{Saffman,Newton}). 
Due to Onsager \cite{Onsager} (for a discussion we refer the
reader to \cite{Sreenivasan}) a statistical treatment of point vortex dynamics is possible 
for equilibrium situations because of the Hamiltonian character of the dynamics, 
provided the ergodic hypothesis holds. Extensions to
non-equilibrium situations based on kinetic equations have been pursued, e.g., by
Joyce and Montgomery \cite{Montgomery}, 
Lundgren and Pointin \cite{Lundgren}, as well as more recently by Chavanis \cite{Chavanis}.   

The purpose of the present article is to generalize Kirchhoff's point vortex model to a rotor 
model that exhibits the formation of large-scale vortical structures due to the formation of 
rotor clusters. 
The existence of such a process in two-dimensional flows where a large-scale vorticity field 
spontaneously emerges from 
an initially random distribution of vortices was first predicted by 
Kraichnan \cite{Kraichnan1} and is termed an inverse cascade. Thereby, the energy that is injected
into the small scales is transfered to larger scales, whereas the enstrophy follows a direct 
cascade from large to small scales. 
It was also Kraichnan \cite{Kraichnan2}, who gave an intuitive explanation of the possible 
mechanism of the cascade: He considered a small-scale axisymmetric vortical structure that 
is exposed to a large-scale strain field. Eventually, the vortex is elongated along the stretching
direction of the strain, i.e. to a first approximation drawn out into an elliptical structure. 
This thinning mechanism induces relative motions between vortices that have been deformed under their
mutual strain, which leads to a decrease of the kinetic energy of the small-scale motion
and consequently to an 
energy transfer upscale. More recently, it has been pointed out numerically and experimentally by Chen 
et al. \cite{Ecke} that the effect 
of vortex thinning is indeed an important feature of the inverse cascade. An appropriate vortex model
for the inverse cascade therefore has to provide a mechanism similar to that identified in  \cite{Ecke}.
 
Although, several point vortex models 
have been known for a long time to form large-scale vortical structures from an initially
random distribution of point vortices due to the events of vortex merging \cite{Benzi,Weiss} or special 
forcing mechanisms \cite{Siggia},
an explicit inclusion of the concept of vortex thinning never has been taken into account.

In our vortex model, the small-scale vortical structure is represented by a rotor consisting of two point vortices 
with equal circulation that are glued together by a nonelastic bond.
The main observation now is that the two co-rotating point vortices mimic a far-field that is similar to an 
elliptical vortex,  which makes the rotor sensitive to a large-scale strain.
The model is motivated by a representation of the vorticity field as a superposition of vortices
with elliptical Gaussian shapes along the lines of Melander, Styczek and Zabusky \cite{Melander}.
The nonelastic bond in a rotor can be considered as an over-damped spring which models the 
influence of forcing and viscous damping. 

However, the main renewal in this 
model is not the mechanism of how the energy is injected into the system, but how the energy is transfered
upscale due to the strain-induced relative motions between the rotors in the sense of vortex thinning.
The efficiency of the cascade in the rotor model is supported by the relatively fast demixing of the system as well
as a Kolmogorov constant of $C_K= 5.89 \pm 0.74$ that is within the range of accepted values 
\cite{Uriel,Yakhot,Bofetta,Tabeling1,Tabeling2}.

This paper is organized as follows: First of all, we consider a decomposition of the vorticity field
into localized vortices with different shapes in section \ref{dec}. In section \ref{ans}, we make an
ansatz for the shapes which corresponds to an elliptical distribution of the vorticity and discuss
the interaction of two vortices with like-signed circulation within the point vortex model, the
Gaussian vortex model and the elliptical model. It will explicitly be shown that the former two
models do not lead to a relative motion between the vortices, and that the thinning mechanism is
only taken into account by the elliptical model. A suitable forcing mechanism for the vorticity equation is 
introduced in section \ref{forcing} and then used within our generalized
vortex model, presented in section \ref{modelsection}.   
\section{Decomposition of the vorticity field into vortices with arbitrary shapes}\label{dec}
As it is known from basic fluid dynamics, the vorticity 
\begin{equation}
 \boldsymbol \omega({\bf x},t)= \nabla \times {\bf u}({\bf x},t)
\end{equation}
only possesses one component in two-dimensional flows and obeys the evolution equation
\begin{equation}\label{omega}
\dot \omega({\bf x},t) + {\bf u}({\bf x},t) \cdot \nabla \omega({\bf x},t) = \nu \nabla^2
\omega({\bf x},t)
\end{equation}
Here, the advecting velocity field is determined by Biot-Savart's law according to
\begin{equation}\label{biot}
 {\bf u}({\bf x},t)= \int \textrm{d} {\bf x}'  \omega({\bf x}',t)  {\bf e}_z \times \frac{{\bf
x}-{\bf x}'}{2 \pi |{\bf x}-{\bf x}'|^2} 
\end{equation}
We consider the two-dimensional vorticity equation in Fourier space 
\begin{equation}\label{vorticity}
\dot \omega({\bf k},t)-i{\bf k} \cdot \int \textrm{d}{\bf k}'  {\bf u}({\bf k}')
 \omega({\bf k}-{\bf k}',t) \omega({\bf k}',t)=-\nu k^2 \omega({\bf
  k},t)
\end{equation}
with ${\bf u}({\bf k})=\frac{i}{4\pi^2}[{\bf e}_z\times \frac{{\bf k}'}
{k'^2}]$. \\ 
In the following the vorticity is decomposed into vortices $\omega_j({\bf k},t)$ with the 
circulation $\Gamma_j$ that are centered at ${\bf x}_j(t)$ and that possess the shapes 
$W_j({\bf k},t)$, namely 
\begin{equation}
 \omega_j({\bf k},t)= \Gamma_j e^{i{\bf k}\cdot {\bf x}_j(t)+W_j({\bf k},t)}
\end{equation}
Our ansatz thus reads
\begin{equation}\label{ansatz}
 \omega({\bf k},t)=\sum_j \omega_j({\bf k},t)=\sum_j \Gamma_j e^{i{\bf k}\cdot {\bf x}_j(t)+W_j({\bf
k},t)}
\end{equation}
For $W_j({\bf k},t)=0$, we recover the vorticity field $\omega({\bf x},t)$ of point vortices
\begin{equation}\label{point}
\omega({\bf x},t)= \sum_j \Gamma_j \delta({\bf x}-{\bf x}_j(t))
\end{equation}
 that are located at the positions ${\bf x}_j(t)$ and that are a solution of the ideal vorticity
equation ($\nu=0$), which conserves the vorticity along a Lagrangian trajectory. Inserting the
vorticity field from (\ref{point}) into Biot-Savart's law (\ref{biot}) immediately yields the
evolution equation for the point vortices
\begin{equation}
 \dot {\bf x}_j(t) = \sum_l \frac{\Gamma_l}{2 \pi} {\bf e}_z \times \frac{{\bf x}_j(t)-{\bf
x}_l(t)}{|{\bf x}_j(t)-{\bf x}_l(t)|^2}
\end{equation}
We now insert our ansatz (\ref{ansatz})
into the vorticity equation and obtain
\begin{eqnarray}\label{evom}\nonumber
&~&\sum_j \Gamma_j e^{i{\bf k}\cdot {\bf x}_j+W_j({\bf k},t)}
\left[ i{\bf k}\cdot \dot {\bf x}_j(t)+\dot W_j({\bf k},t) +\nu k^2
\right]\\ \nonumber
 &=&
i{\bf k} \cdot \sum_{j,l}\Gamma_j \Gamma_l
\int \textrm{d}{\bf k}'  {\bf u}({\bf k}')
e^{i({\bf k}-{\bf k}')\cdot {\bf x_j}+i{\bf k}'\cdot {\bf x}_l} \\
&~&\times e^{W
_j({\bf k}-{\bf k}',t)+W_l({\bf k}',t)}
\end{eqnarray}
The left-hand side of this equation contains the sweeping dynamic of the vortices 
encoded in the temporal change of ${\bf x}_j(t)$ as well as the temporal change of the shapes
$W_j({\bf k},t)$ due to shearing and vorticity. In the inviscid case, the entire dynamic of the
$j$-th vortex is determined by the nonlinearity on the right hand side of equation (\ref{evom})
which couples the different Fourier modes of the vortices $l$ as well as the self-interaction term
from $j=l$ in a rather complicated manner.
  
Nevertheless, a separation of the effects becomes possible under the assumption that the overlap of
the different vortex structures is negligible, which is valid for widely separated vortices. To this
end, we single out the terms in the summations over $j$ and get
\begin{eqnarray}\label{evom_neu}\nonumber
&~& i{\bf k}\cdot \dot {\bf x}_j(t)+\dot W_j({\bf k},t) +\nu k^2\\ \nonumber
 &=& i{\bf k} \cdot \sum_{l} \Gamma_l
\int \textrm{d}{\bf k}'  {\bf u}({\bf k}')
e^{-i{\bf k}'\cdot ({\bf x_j}- {\bf x}_l)}\\ 
&~&\times e^{W_j({\bf k}-{\bf k}',t)-W_j({\bf k},t)+W_l({\bf k}',t)}
\end{eqnarray}
The sweeping dynamic can now be defined via the terms in the evolution equation (\ref{evom_neu})
which are 
proportional to ${\bf k}$. This immediately yields the evolution equations for the center of the
vortices
\begin{equation}\label{x_j}
\dot {\bf x}_j(t)=\sum_{l} \Gamma_l
{\bf U}_{jl}({\bf x}_j-{\bf x}_l)
\end{equation}
where we have defined the velocity kernels
\begin{equation}\label{U}
{\bf U}_{jl}({\bf r})=
\int \textrm{d}{\bf k}' {\bf u}({\bf k}')e^{-i{\bf k}'\cdot {\bf r}}
e^{W_j(-{\bf k}',t)+W_l({\bf k}',t)}
\end{equation}
Inserting the evolution equation of the vortex centers back into \eqref{evom} yields the evolution
equations for the shapes
\begin{eqnarray}\label{shape}\nonumber
\lefteqn{
\dot W_i({\bf k},t)=-\nu k^2 }
\\
&&+i
{\bf k}\cdot \sum_{l} \Gamma_l 
\int d{\bf k}' {\bf u}({\bf k}')
e^{-i{\bf k}'\cdot [{\bf x}_i-{\bf x}_l]}
e^{W_i(-{\bf k}',t)+W_l({\bf k}',t)} 
\nonumber 
\\
&&\times
\left[
e^{W_i({\bf k}-{\bf k}',t)-W_i(-{\bf k}',t)-W_i({\bf k},t)}-1\right]
\end{eqnarray}
Here the sum includes also the self-interaction term with $j=l$.
The system of equations (\ref{x_j}) and (\ref{shape}) is the extension of the set of evolution equations
for the
$\delta$-point vortices (\ref{point}) and takes into account possible changes of the shapes
$W_j({\bf k},t)$ of
each vortex.
It is important to stress that up to now we did not impose any restrictions on the shapes $W_j({\bf
k},t)$. 

\section{Approximation via vortices with elliptical shapes}\label{ans}
The vorticity of an elliptical vortex with major and minor semi-axes ${\bf a}$ and ${\bf b}$ can be
written according to
\begin{equation}\label{elli}
 \omega({\bf x},t)= \frac{\Gamma}{ \pi} e^{-{\bf x} C {\bf x}}
\end{equation}
where $C={\bf a} {\bf a} +{\bf b}{\bf b}$ is the symmetric matrix of the dyadic products of the
semi axes. A rotation of the coordinate system then turns (\ref{elli}) into  
\begin{equation}
 \omega({\bf x},t)= \frac{\Gamma}{ \pi} e^{-\left(q^{-2} x^2 + q^2 y^2 \right)}  
\end{equation}
where $q$ is the is the ratio of the major to the minor semi axes. The vorticity in Fourier space
thus reads
\begin{equation}
\omega({\bf k},t)= \Gamma e^{-\left (q^{2} k_x^2 + q^{-2} k_y^2 \right)}
\end{equation}
which again corresponds to an elliptical distribution of the vorticity. An elliptical representation
of the shapes can thus be obtained via the  
approximation
\begin{equation}
W_j({\bf k},t)\approx -\frac{1}{2} {\bf k} C_j(t) {\bf k}
\end{equation}
with the symmetric matrix 
$C_j(t)$. In approximating the last term on the right-hand
side of equation (\ref{shape}) by
\begin{eqnarray}\nonumber
&~&e^{W_j({\bf k}-{\bf k}',t)-W_j(-{\bf k}',t)-W_j({\bf k},t)}-1\\
&\approx& 
-\frac{1}{2}[{\bf k} C_j(t){\bf k}'+{\bf k}' C_j(t) {\bf k}]
\end{eqnarray}
we are able to derive an evolution equation for the matrix $C_j(t)$, namely 
\begin{eqnarray}\label{evolC}
\dot C_j&=&2\nu E+
\Gamma_j [S_{jj} C_j+C_j S_{jj}^T]\\ \nonumber
&~&+
\sum_{l \ne j} \Gamma_l [S_{jl}({\bf x}_j-{\bf x}_l)
C_j+C_j S_{jl}({\bf x}_j-{\bf x}_l)^T]
\end{eqnarray}
Here, we explicitly have introduced
the matrix $S_{jl}=[\nabla {\bf U}_{jl}({\bf x}_j-{\bf x}_l)]$
and have singled out the term with $j=l$.
The velocity field is now determined from Eq. (\ref{U}) up to the first
order in $C_j+C_l$ valid for widely separated vortices
\begin{eqnarray}
{\bf U}_{jl}({\bf r}) &=& 
\int \textrm{d}{\bf k}' \cdot {\bf u}({\bf k}') e^{-i{\bf k}'\cdot {\bf r}}
e^{-\frac{1}{2} {\bf k}'(C_j+C_l){\bf k}'}
\nonumber \\
&\approx &
{\bf e}_z \times \left[1+\frac{1}{2}
\nabla_{\bf r} (C_j+C_l) \nabla_{\bf r} \right]
\frac{{\bf r}}{2\pi|{\bf r}|^2}
\end{eqnarray}
The evolution equation for the vortex centers then reads
\begin{eqnarray}\label{x_i}
\dot {\bf x}_j &=& \sum_l \Gamma_l {\bf e}_z \times \frac{{\bf x}_j-{\bf x}_l}
{2\pi|{\bf x}_j-{\bf x}_l|^2}\\ \nonumber
&~&+\sum_l \Gamma_l \nabla_{{\bf x}_j}[C_j+C_l]\nabla_{{\bf x}_j}
{\bf e}_z \times
\frac{{\bf x}_j-{\bf x}_l}
{4\pi|{\bf x}_j-{\bf x}_l|^2}
\end{eqnarray}
A similar system of equations (\ref{evolC}) and (\ref{x_i}) has been obtained by Melander et al.
\cite{Melander} by means of a truncation of the stream function within their
second-order moment model for the Euler equations.

\section{Motion of vortices with equal circulation within the different models}\label{models}
It is illustrative to consider the interaction of two vortices $1$ and $2$ at the positions ${\bf
x}_1$ and ${\bf x}_2$ that possess equal circulation
$\Gamma_1=\Gamma_2=\Gamma$ in the realm of the different vortex models considered above, namely the
point vortex model, the Gaussian shape model, and the elliptical Gaussian shape model. \\
~\\


{\em i.) Gaussian shapes:}\\

Let us consider the case of Gaussian shapes $C_1= c_1E$ and $C_2= c_2 E$.
The symmetry of the problem imposes that $c_1=c_2=c$, and we arrive at the
following evolution equations for the centers
\begin{eqnarray}\nonumber
 \dot {\bf x}_1&=& \Gamma \int \textrm{d}{\bf k}' {\bf u}({\bf k}') e^{-i {\bf k}' \cdot ({\bf
x}_1-{\bf x}_2)} e^{-\frac{1}{2} c k'^2}\\ 
 \dot {\bf x}_2&=& \Gamma \int \textrm{d}{\bf k}' {\bf u}({\bf k}') e^{-i {\bf k}' \cdot ({\bf
x}_2-{\bf x}_1)} e^{-\frac{1}{2} c k'^2}
\end{eqnarray}
In making use of
\begin{eqnarray} \nonumber
&~& \frac{i}{4 \pi^2}{\bf e}_z \times \int \textrm{d} {\bf k}' \frac{{\bf k}'}{k'^2} e^{-i{\bf
k}'\cdot {\bf r}}e^{-\frac{1}{2} c k'^2}\\ \nonumber
&=&-\frac{1}{4 \pi^2} {\bf e}_z \times \nabla_{\bf r}
(\nabla_{\bf r}^2)^{-1} 
\int \textrm{d} {\bf k}' e^{-i{\bf
k}'\cdot {\bf r}} e^{-\frac{1}{2}c k'^2} \\ \nonumber
&=& -\frac{1}{2 \pi} {\bf e}_z \times \nabla_{\bf r}
(\nabla_{\bf
r}^2)^{-1} \frac{\Gamma}{2 \pi c} e^{-\frac{r^2}{2 c}}\\ 
&=& \frac{\Gamma}{2 \pi}\Big(1- e^{-\frac{r^2}{2
c}} \Big) {\bf e}_z \times \frac{{\bf r}}{r^2}
\end{eqnarray}
which is the velocity profile of a Lamb-Oseen vortex, the evolution equation for the 
relative coordinate reads
\begin{equation}
 \dot {\bf r}=\frac{\Gamma}{ \pi}\Big(1- e^{-\frac{r^2}{2
c}}\Big) {\bf e}_z \times \frac{{\bf r}}{r^2}
\end{equation}
The evolution equation for the shapes $c$ has to be evaluated in a similar fashion from Eq. 
(\ref{evolC}), but for now
we invoke the approximation
\begin{equation}
 \dot c= 2 \nu
\end{equation}
where we have neglected the interaction-terms in (\ref{evolC}. 
This yields the evolution equations for two Lamb-Oseen vortices 
\begin{eqnarray} 
 \dot r&=&0 \\
 \dot \varphi &=& \frac{\Gamma}{\pi r^2} \Big(1-e^{-\frac{r^2}{4 \nu t}}\Big) \label{Gauss}
\end{eqnarray}
In the case of vanishing viscosity, we recover the evolution equations of a point vortex pair
which undergoes a circular motion around the center ${\bf R}=\frac{{\bf x}_1+{\bf
x}_2}{2}$
with the angular velocity $ \frac{\Gamma}{\pi r^2}$.
Compared to that case, the angular velocity of
the Gaussian vortex patches described by Eq. (\ref{Gauss}) is thus slowed down by viscosity. 

However, if we observe such two vortices in real flows, we would see a deformation of the two vortices 
due to their mutual strain. This deformation in turn, leads to an attractive motion of the vortex centers 
and ultimately to a merging process of the two vortices.  
At this point, it is important to notice that a direct consequence of an axisymmetric vorticity 
profile is that $\dot r=0$, which means that no relative motion is induced. 
Furthermore, in this context we want to mention that a recent investigation of the two-point vorticity statistic
in two-dimensional turbulence within a Gaussian approximation revealed the absence of an energy flux from 
smaller scales to larger scales \cite{arXiv}. The emergence of deformable structures that induce such relative 
motions in the context of vortex thinning can thus be considered as an important feature of the inverse cascade.\\
~\\   
{\em ii.) Elliptical shapes:}\\

As we have discussed in {\em i.)}, the mutual interaction of Gaussian vortices in real flows leads 
to deformations and subsequently attractive motions of the vortex centers. Such deformations
can be considered in a first approximation as elliptical deformations.\\
Therefore, the interaction of two elliptical vortices should for the first time lead to non-vanishing 
relative motions.\\
The evolution equation for two elliptically shaped vortices read
\begin{eqnarray}\nonumber
 \dot {\bf x}_1&=& \Gamma \int \textrm{d}{\bf k}' {\bf u}({\bf k}') e^{-i {\bf k}' \cdot ({\bf
x}_1-{\bf x}_2)} e^{-\frac{1}{2} {\bf k}' [C_1+C_2] {\bf k}'}\\ 
 \dot {\bf x}_2&=& \Gamma \int \textrm{d}{\bf k}' {\bf u}({\bf k}') e^{-i {\bf k}' \cdot ({\bf
x}_2-{\bf x}_1)} e^{-\frac{1}{2} {\bf k}' [C_1+C_2] {\bf k}'}
\end{eqnarray}
For widely separated vortices the evolution equation for the relative coordinate thus reads 
\begin{eqnarray}
\dot {\bf r} =  \frac{\Gamma}{ \pi}\left(1 +\frac{1}{2} \nabla_{{\bf r}}[C_1+C_2]\nabla_{{\bf r}}
\right) {\bf e}_z \times \frac{{\bf r}}
{r^2}
\end{eqnarray}
which can lead to contributions to the relative motion $\dot r \neq 0$, provided that the matrices
$C_1$ and $C_2$ do not reduce to diagonal matrices as in the case of Gaussian shapes. Whether the motion
is attractive or repulsive, is to a far extend determined by the alignment angle $\varphi_a-
\varphi_r$ between ${\bf r}$ and the major semi axis ${\bf a}$ of the vortices, which is explicitly
derived for the interaction of two rotors in section \ref{inter}, for instance in Eq.
(\ref{R_i}).

\section{The forcing mechanism}\label{forcing}
As it can be seen from Eq. (\ref{shape}), the viscous contributions causes the broadening of
the shape of a vortex. Since this effect is more pronounced for smaller vortex structures, thus 
larger values of $k^2$ in (\ref{evolC}), an 
appropriate forcing mechanism has to counteract this effect and provide an energy input at small scales.
The forcing mechanism we want to introduce, consists in forcing the semi axes of each elliptical vortex and thus 
the whole shape of this vortex back to a fixed shape $C_0$. It will be seen in section
\ref{modelsection} that the influence of this kind of forcing makes the two like-signed
point vortices of our rotor model to behave as if they were connected by an over-damped spring.

The described forcing mechanism can now be introduced in the
following way:
\begin{eqnarray}\label{modelg1}
\dot \Gamma_i &=&-a\Gamma_i +f_i
 \\
\dot {\bf x}_i &=& \sum_{l} \Gamma_l
{\bf U}_{il}({\bf x}_i-{\bf x}_l)+{\bf U}_i(t)
 \\ \nonumber
\dot C_i&=&2 \nu E+\gamma(C_0-C_i)+
\Gamma_i [S_{ii} C_i+C_i S_{ii}^T]
\nonumber \\ \label{modelg3}
&+&
\sum_l \Gamma_l [S_{il}({\bf x}_i-{\bf x}_l)
C_i+C_i S_{il}({\bf x}_i-{\bf x}_l)^T]
\end{eqnarray}
Such type of forcing may be obtained from the vorticity equation
(\ref{vorticity}) by just adding a linear damping term,
$-a\omega({\bf k},t)$ as well as the forcing term $F({\bf k},t)$,
\begin{eqnarray}
F({\bf k},t) &=&\label{force}
\sum_j f_j e^{i {\bf k}\cdot \tilde {\bf x}_j(t)+\tilde W_j}
\approx
\sum_j e^{i {\bf k}\cdot {\bf x}_j(t)+W_j}
\\
&\times&
[f_j + i \Gamma_j {\bf k}\cdot {\bf U}_j(t)-
\frac{1}{2}\Gamma_j {\bf k} \gamma \left( C_0(t)- C_j(t)\right) {\bf k}]
\nonumber 
\end{eqnarray}
where the centers $\tilde {\bf x}_j= {\bf x}_j(t)+\frac{\Gamma_j}{f_j} {\bf U}_j(t)$ 
as well as the shapes $\tilde W_j = W_j -
\frac{\gamma}{2}\frac{\Gamma_j}{f_j}
{\bf k} \left( C_0(t)- C_j(t)\right) {\bf k}
$ are close to the centers and the shapes of
the elliptical vortices.

The first contribution in Eq. (\ref{force})
leads to a modulation of the circulation, the second
term describes a shift of the rotor center and the third one corresponds to
a modification of the width of the Gaussian vortex shape that forces the elliptical vortex back to a
certain shape $C_0$. The stretching of the semi axes of the elliptical vortex due to viscous broadening
represented by the first term on the right-hand side in 
Eq. \ref{modelg3} is thus counteracted by the second term trying to contract the shape of the vortex 
back to $C_0$.

A striking analogy to this forcing mechanism can be found
in the explanation of the magneto-rotational instability
\cite{Balbus}. Thereby, two elements of an electrically conduct-
ing fluid that undergo a rotation around a fixed center
are supposed to be connected by an elastic spring repre-
senting the magnetic field. As a consequence the angular
momentum of the system is not a conserved quantity
anymore and the fluid motion becomes unstable.

Although, the introduced forcing mechanism is an ad-hoc forcing, it emerges in a physically plausible way 
from the basic equations of the elliptical model \ref{evolC} and \ref{x_i}.
Furthermore, it should be mentioned that the system of equations (\ref{modelg1}) can be obtained from the
Instanton equations of two-dimensional turbulence by means of a variational ansatz with Gaussian
elliptical vortices \cite{Kolja}.     

\section{Formulation of the rotor model}\label{modelsection}

As we have seen in section \ref{models} about the interaction between two point vortices with equal
circulation compared to the interaction between two elliptical vortices with equal circulation, the
former model fails to describe a relative motion $\dot r$ in the direction of ${\bf r}$. The
thinning mechanism mentioned in \cite{Ecke} is thus clearly neither captured by Onsager's point
vortex model nor by a Gaussian distribution of the vorticity, in analogy to \cite{arXiv}.

\begin{figure}[h]
 \includegraphics[width= 0.5 \textwidth]{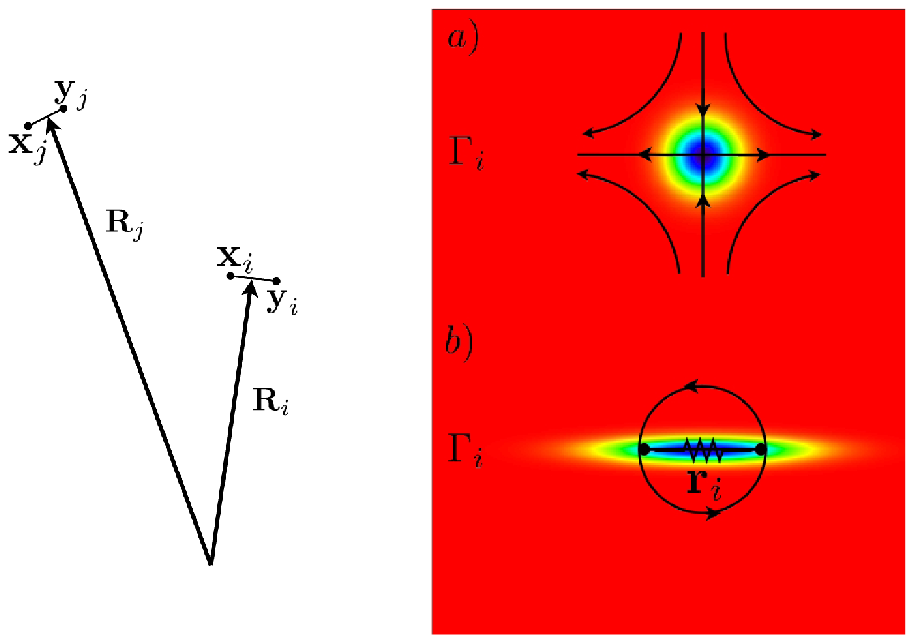}
\caption{(Color online) left: Interaction of two rotors ${\bf R}_i$ and ${\bf R}_j$ with circulation $\Gamma_i$ and
$\Gamma_j$. A rotor is composed of two vortices ${\bf x}_i$ and ${\bf y}_i$ with equal circulation
$\Gamma_i$. For the rotor centered at ${\bf R}_j$, the circular motion of ${\bf x}_i$ and ${\bf
y}_i$ around their center coordinate ${\bf R}_i$ mimics an infitely-thin elliptical vortex with
semi axes ${\bf r}_i={\bf x}_i - {\bf y}_i$, which can be seen from the similarity between the 
multipole expansion (\ref{locR}) and Eq. (\ref{x_i}). This leads to a relative motion between 
the two rotors $\dot R_{ij}=\frac{\textrm{d}}{\textrm{d}t} |{\bf R}_i-{\bf R}_j|$, 
which is not present in the point
vortex model and thus can be considered as an important fingerprint of the inverse cascade of
two-dimensional turbulence.\\
right: a) The thinning mechanism of a circular vortex with circulation $\Gamma_i$ in a shear
velocity field. The rotor in b) is sensitive to such kind of shearing due to his special elliptical
shape. It is sheared in the direction of the semi axes ${\bf r}_i$ and since the elongated structure
is exposed to dissipation the vortex tends to broaden away. The forcing mechanism now forces the
vortex back to a certain structure, acting as an overdamped spring between the two like-signed
vortices of one rotor.}
\label{vector}
\end{figure}
\begin{figure*}
\begin{minipage}[l]{0.325 \textwidth}
\centering
\includegraphics[width=1 \textwidth]{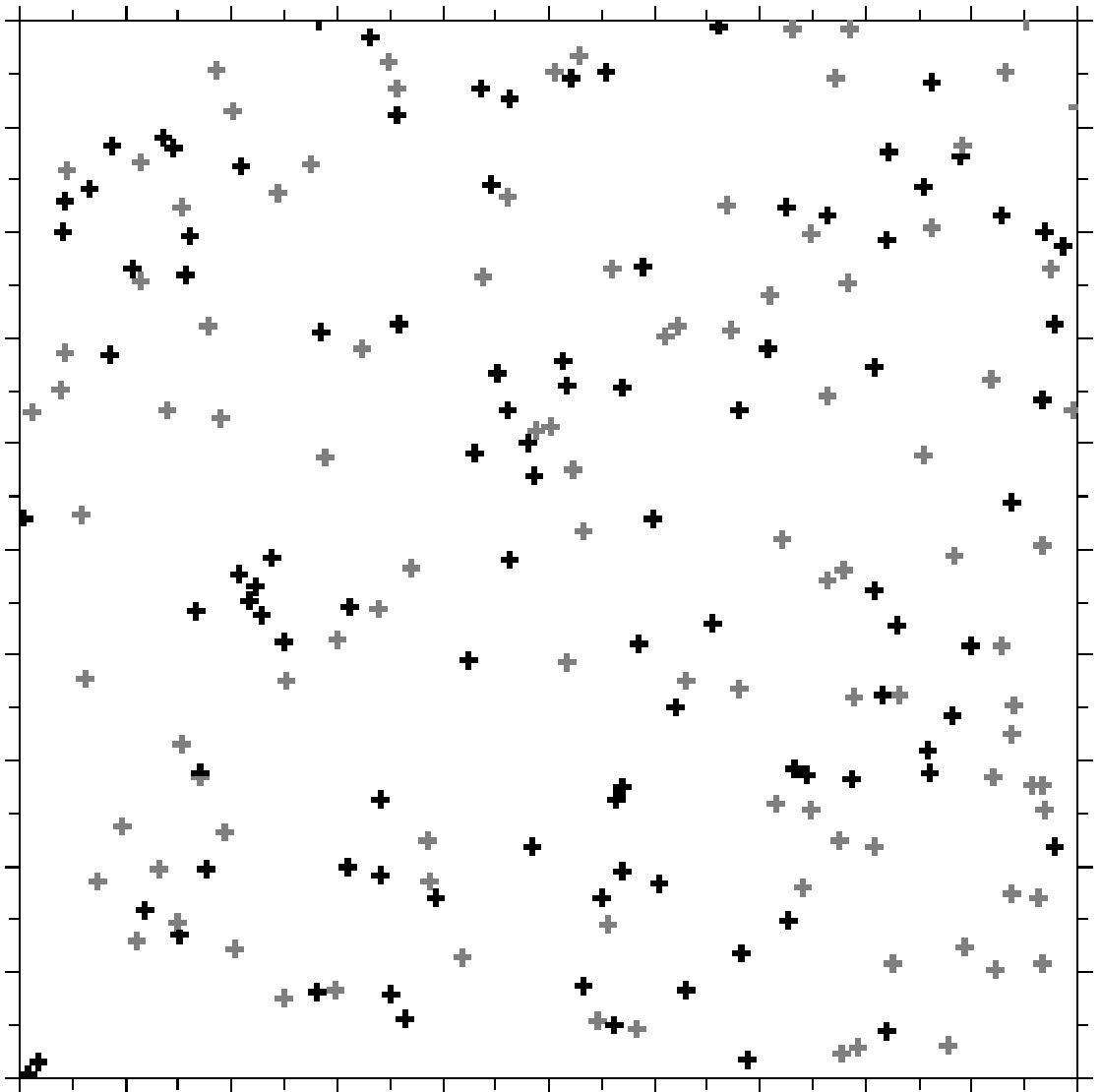}
\small{$t=0$}
\end{minipage}
\begin{minipage}[l]{0.325 \textwidth}
\centering
\includegraphics[width=1 \textwidth]{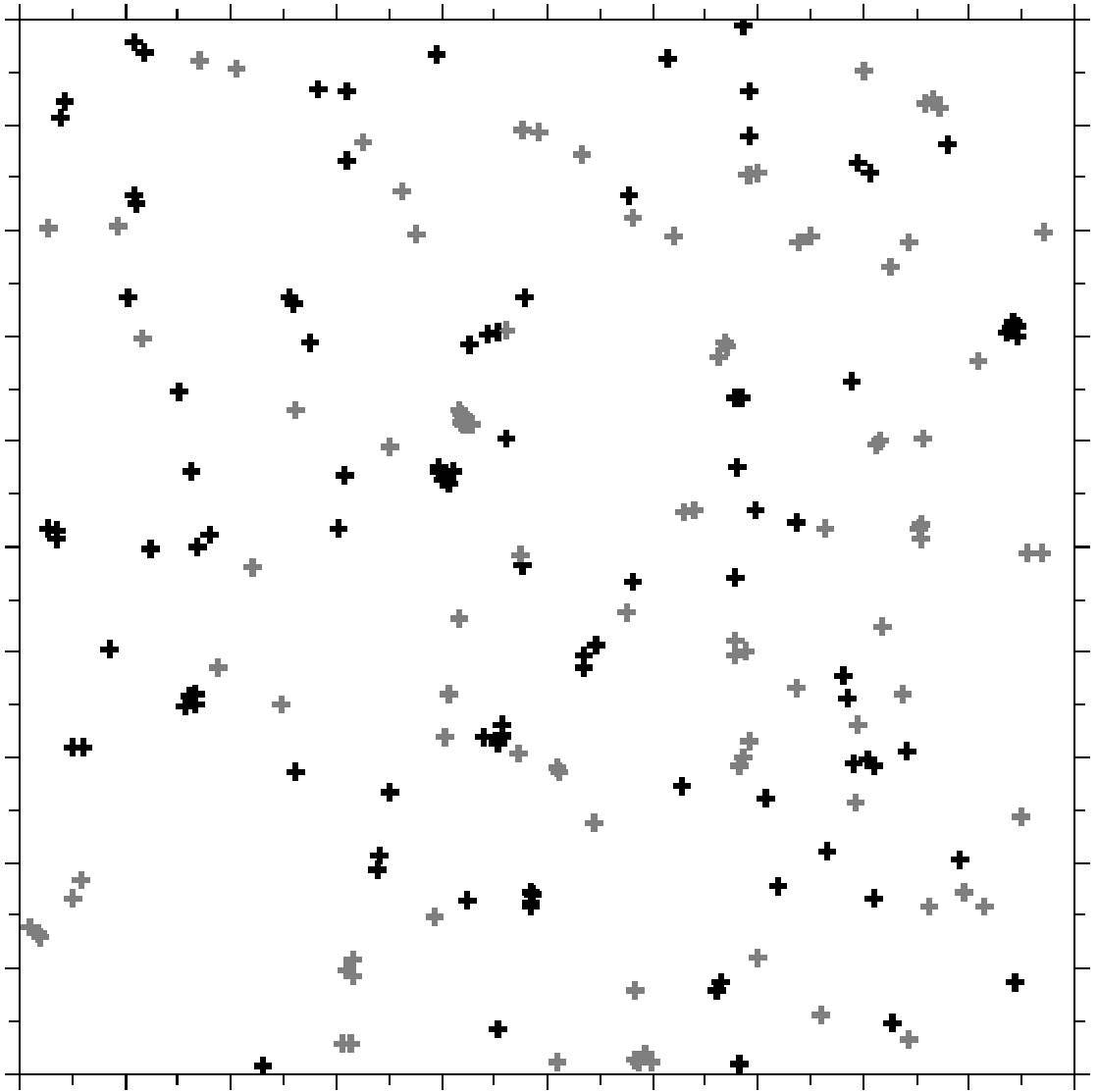}
\small{$t=78.54 T_R$}
\end{minipage}
\begin{minipage}[l]{0.325 \textwidth}
\centering
\includegraphics[width=1 \textwidth]{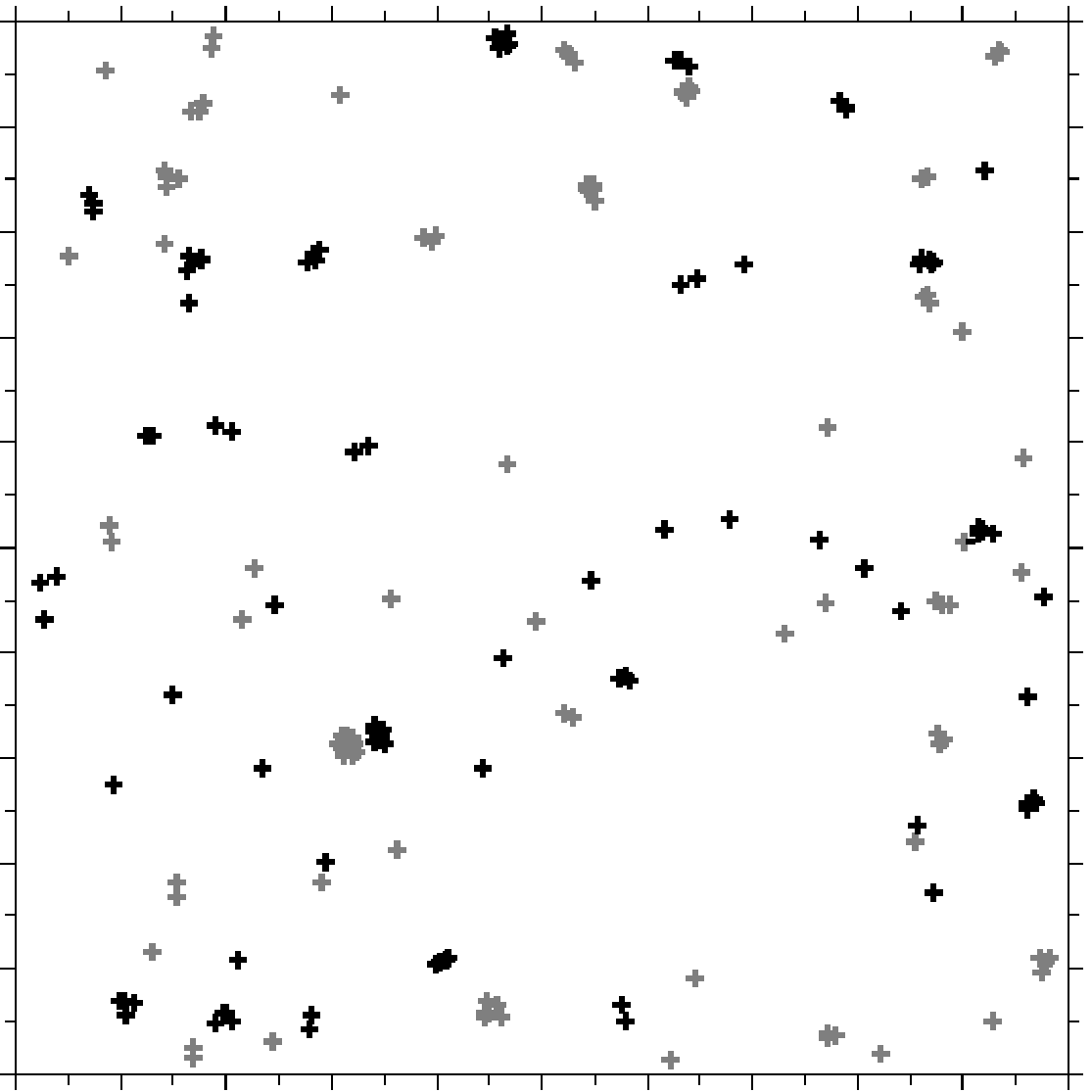}
\small{$t=157.08 T_R$}
\end{minipage}
\begin{minipage}[l]{0.325 \textwidth}
\centering
\includegraphics[width=1 \textwidth]{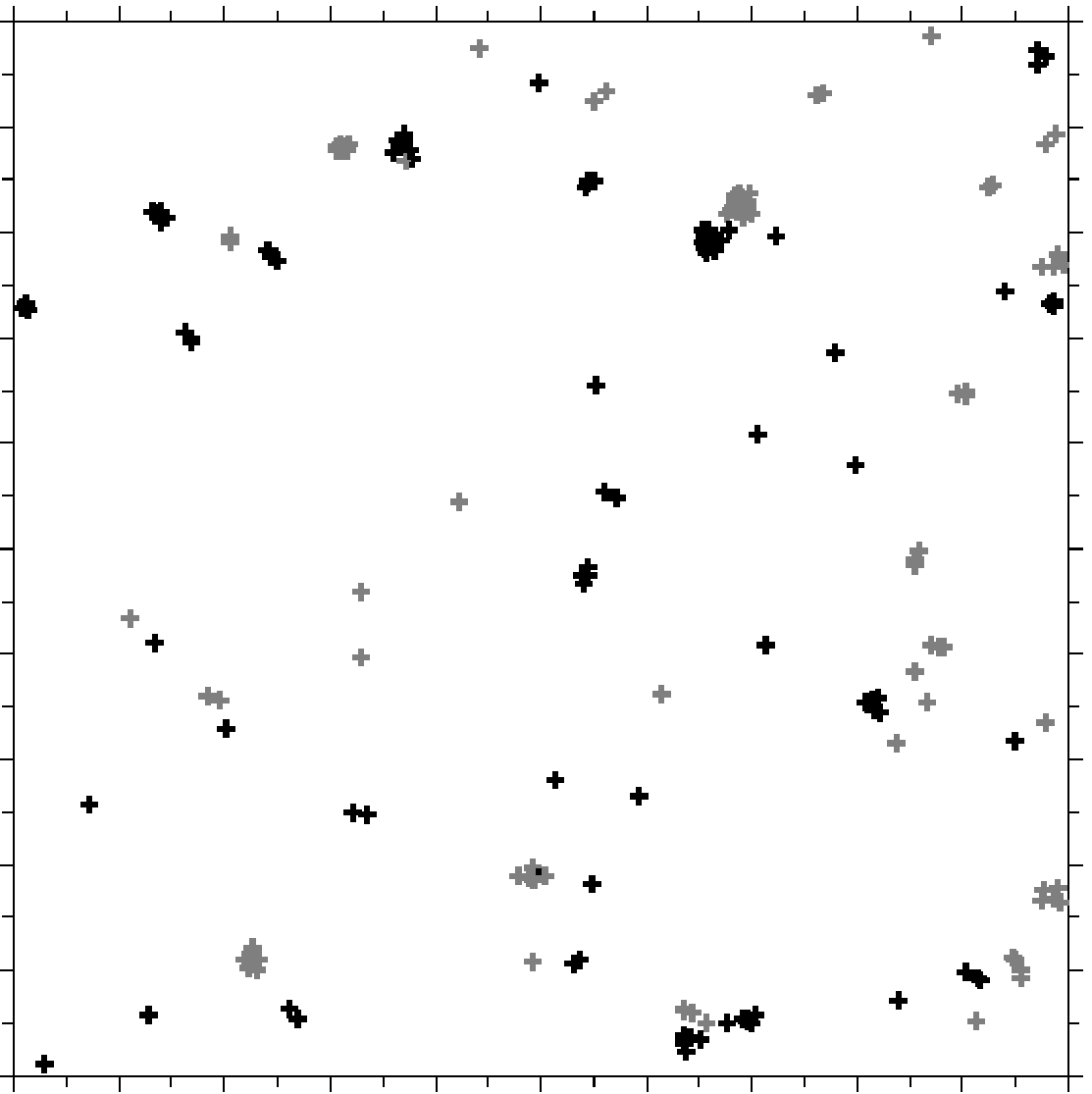}
\small{$t=314.17 T_R$}
\end{minipage}
\begin{minipage}[l]{0.325 \textwidth}
\centering
\includegraphics[width=1 \textwidth]{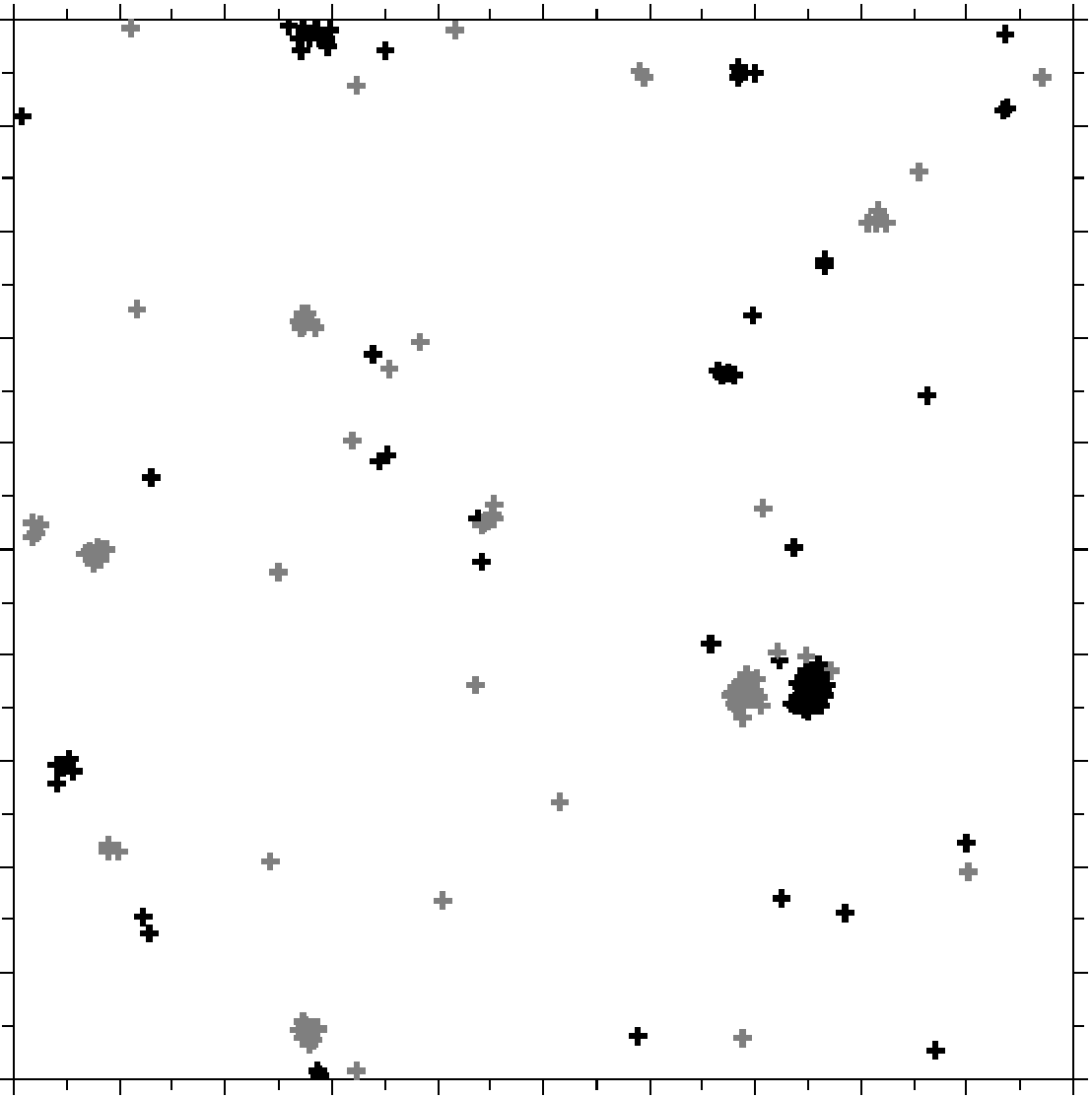}
\small{$t=785.42 T_R$}
\end{minipage}
\begin{minipage}[l]{0.325 \textwidth}
\centering
\includegraphics[width=1 \textwidth]{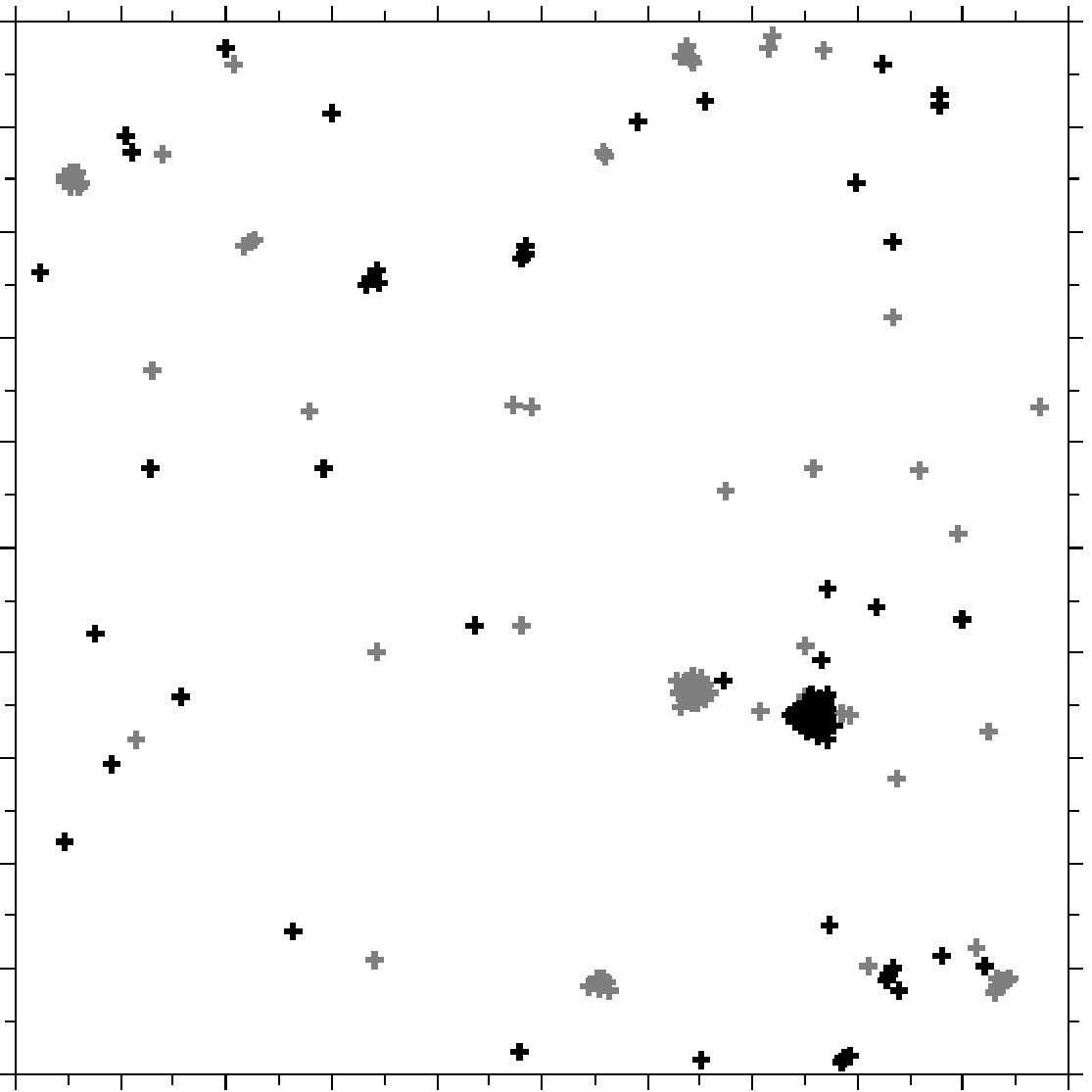}
\small{$t=1233.11 T_R$}
\end{minipage}
\caption{Inverse cascade within the rotor model. Starting from 200 randomly distributed rotors with
circulations $\Gamma= \pm 2 \pi$, rotor clusters of like-signed circulation begin to form. Eventually, a
dipole-cluster is formed that moves through the box attracting the remaining vortices.}
\label{unequal}
\end{figure*} 

\begin{figure*}[t]
\begin{minipage}[l]{0.325 \textwidth}
\centering
\includegraphics[width=1 \textwidth]{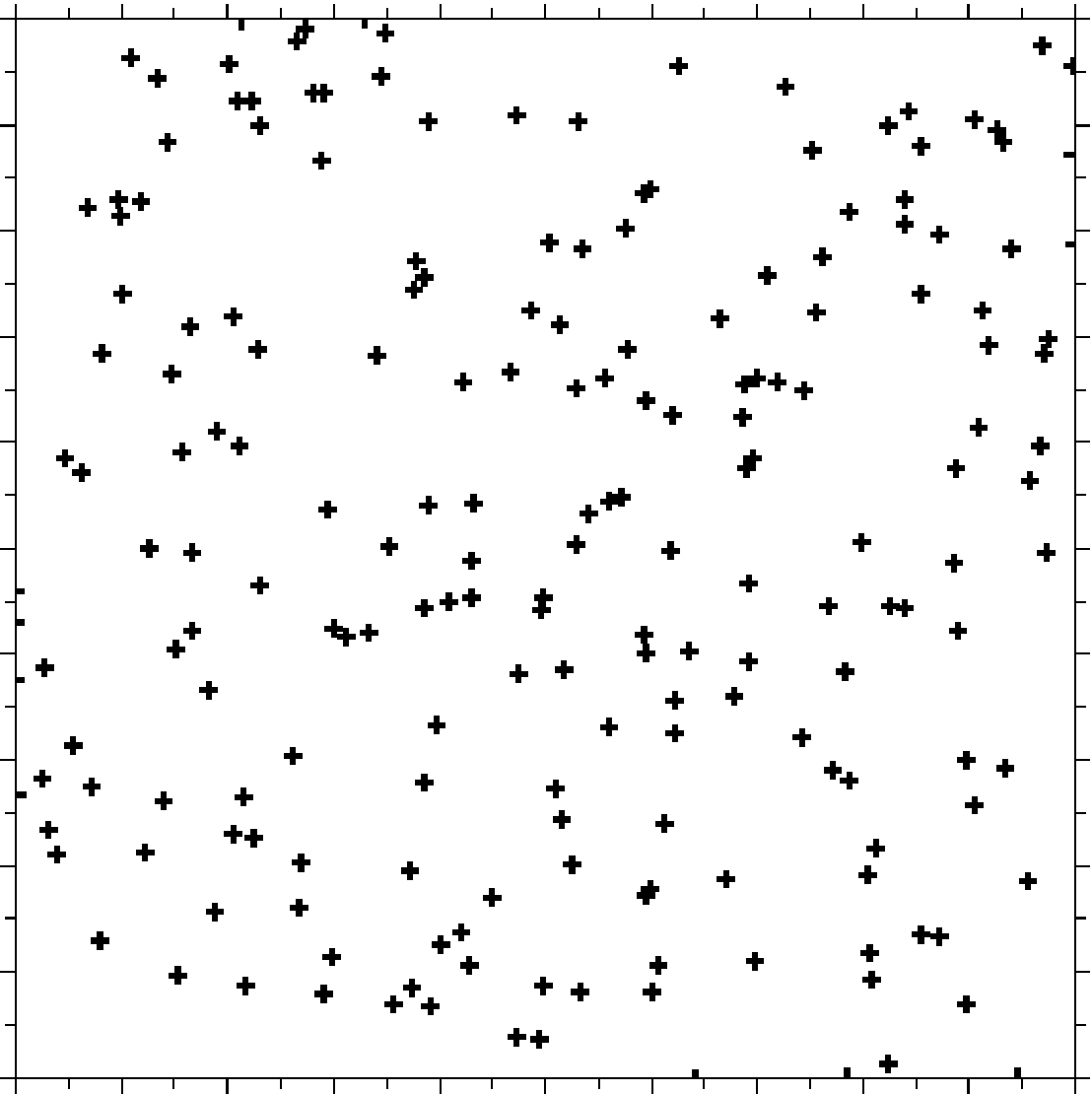}
\small{$t=0$}
\end{minipage}
\begin{minipage}[l]{0.325 \textwidth}
\centering
\includegraphics[width=1 \textwidth]{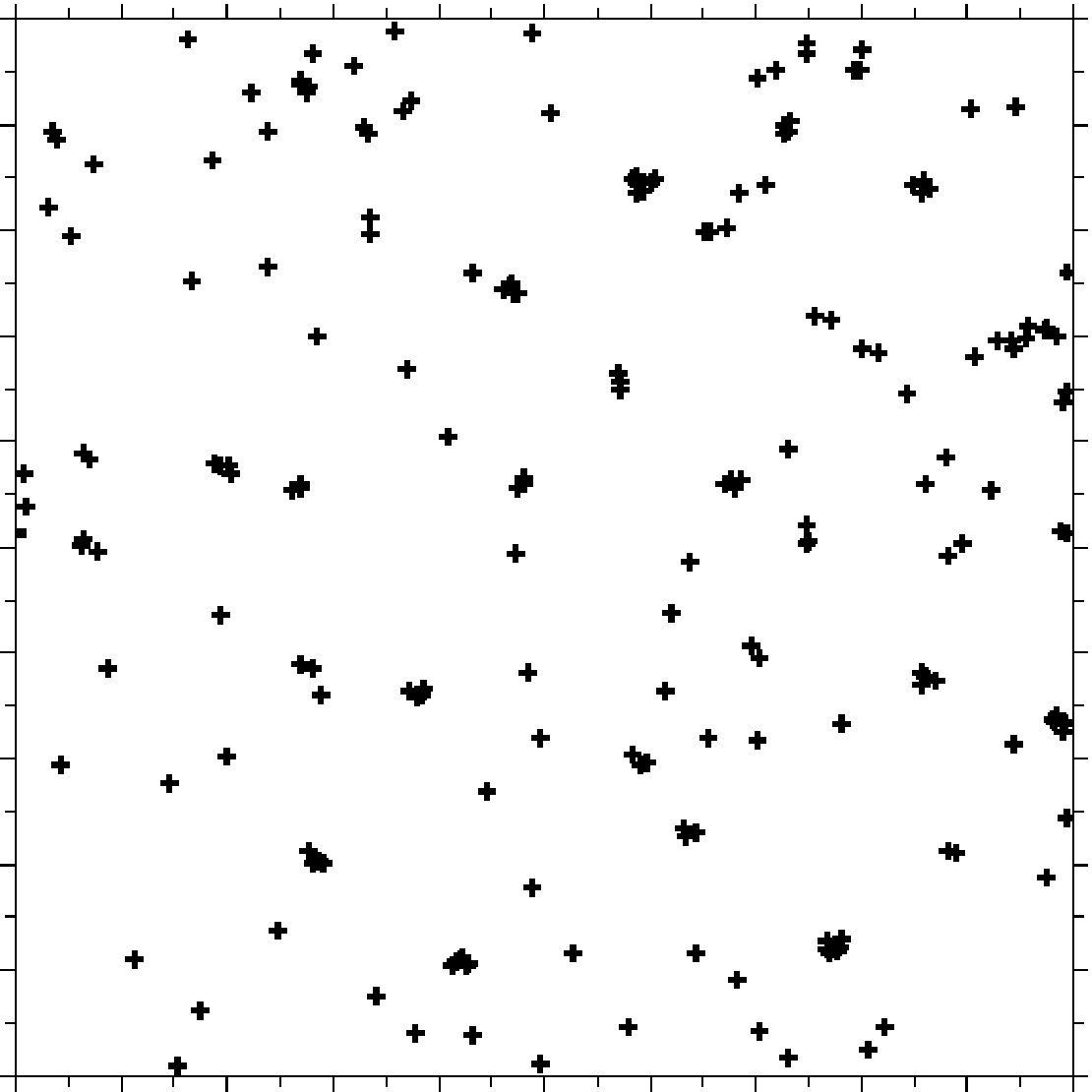}
\small{$t=78.54 T_R$}
\end{minipage}
\begin{minipage}[l]{0.325 \textwidth}
\centering
\includegraphics[width=1 \textwidth]{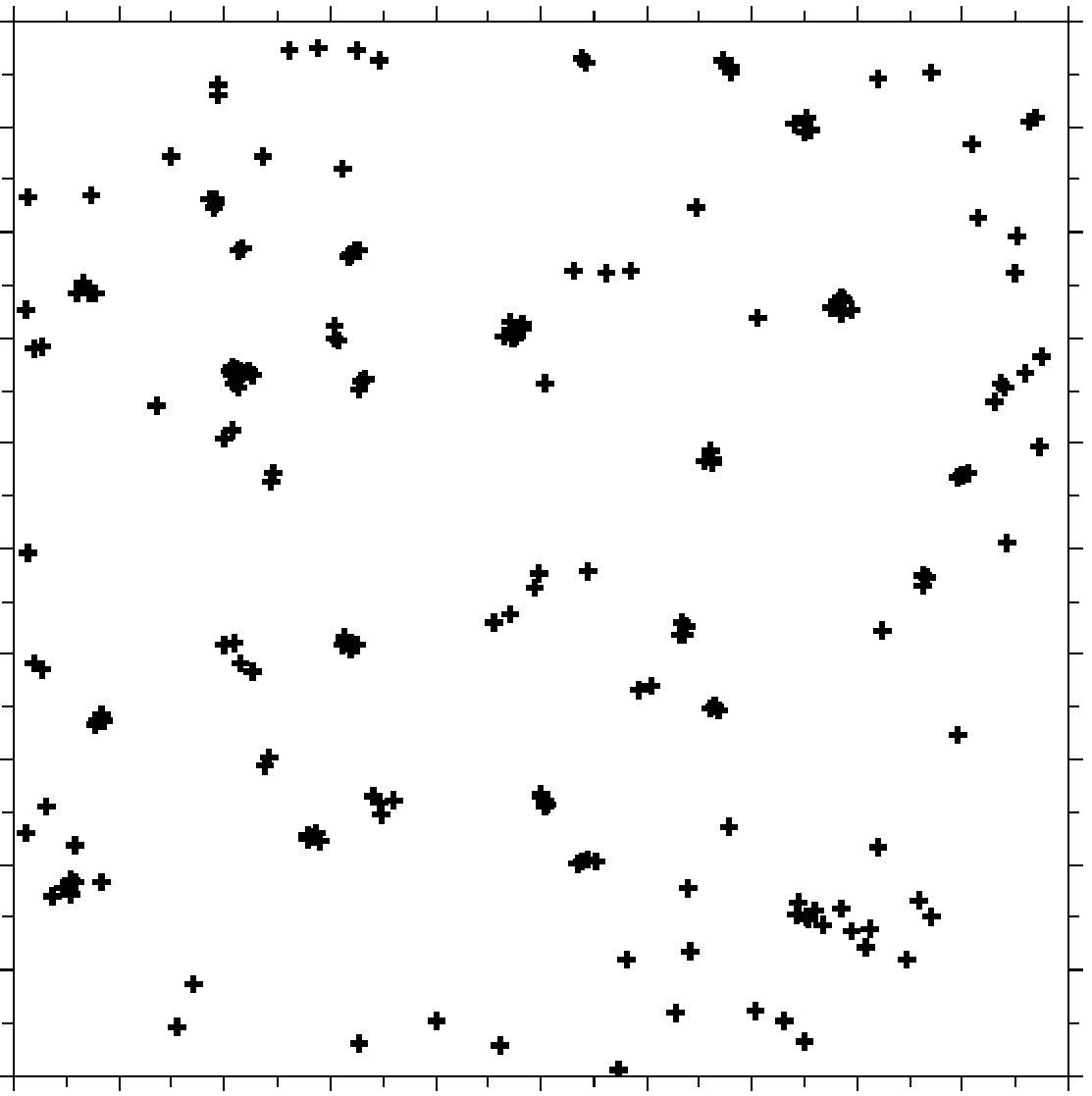}
\small{$t=157.08 T_R$}
\end{minipage}
\begin{minipage}[l]{0.325 \textwidth}
\centering
\includegraphics[width=1 \textwidth]{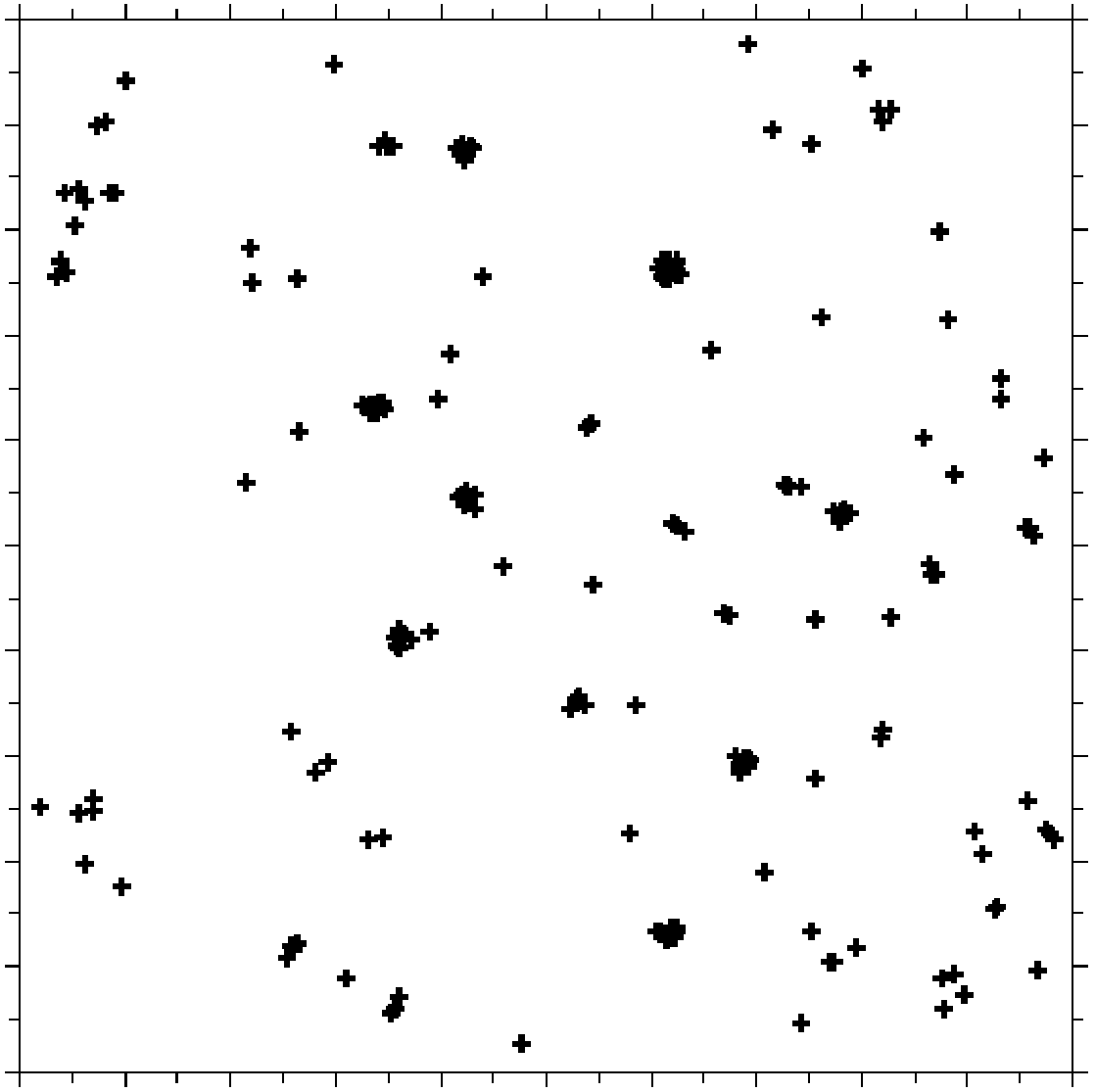}
\small{$t=314.17 T_R$}
\end{minipage}
\begin{minipage}[l]{0.325 \textwidth}
\centering
\includegraphics[width=1 \textwidth]{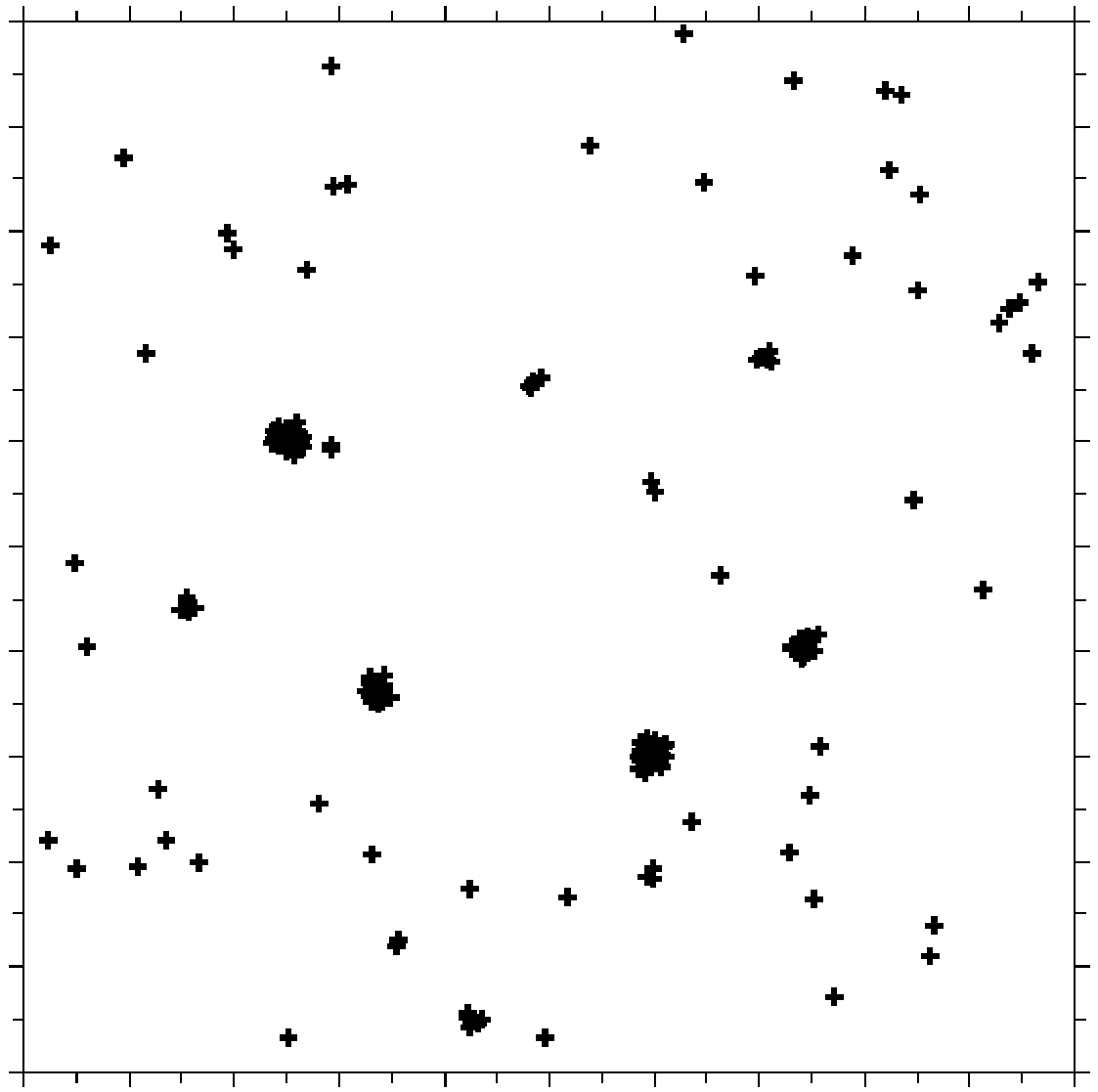}
\small{$t=785.42 T_R$}
\end{minipage}
\begin{minipage}[l]{0.325 \textwidth}
\centering
\includegraphics[width=1 \textwidth]{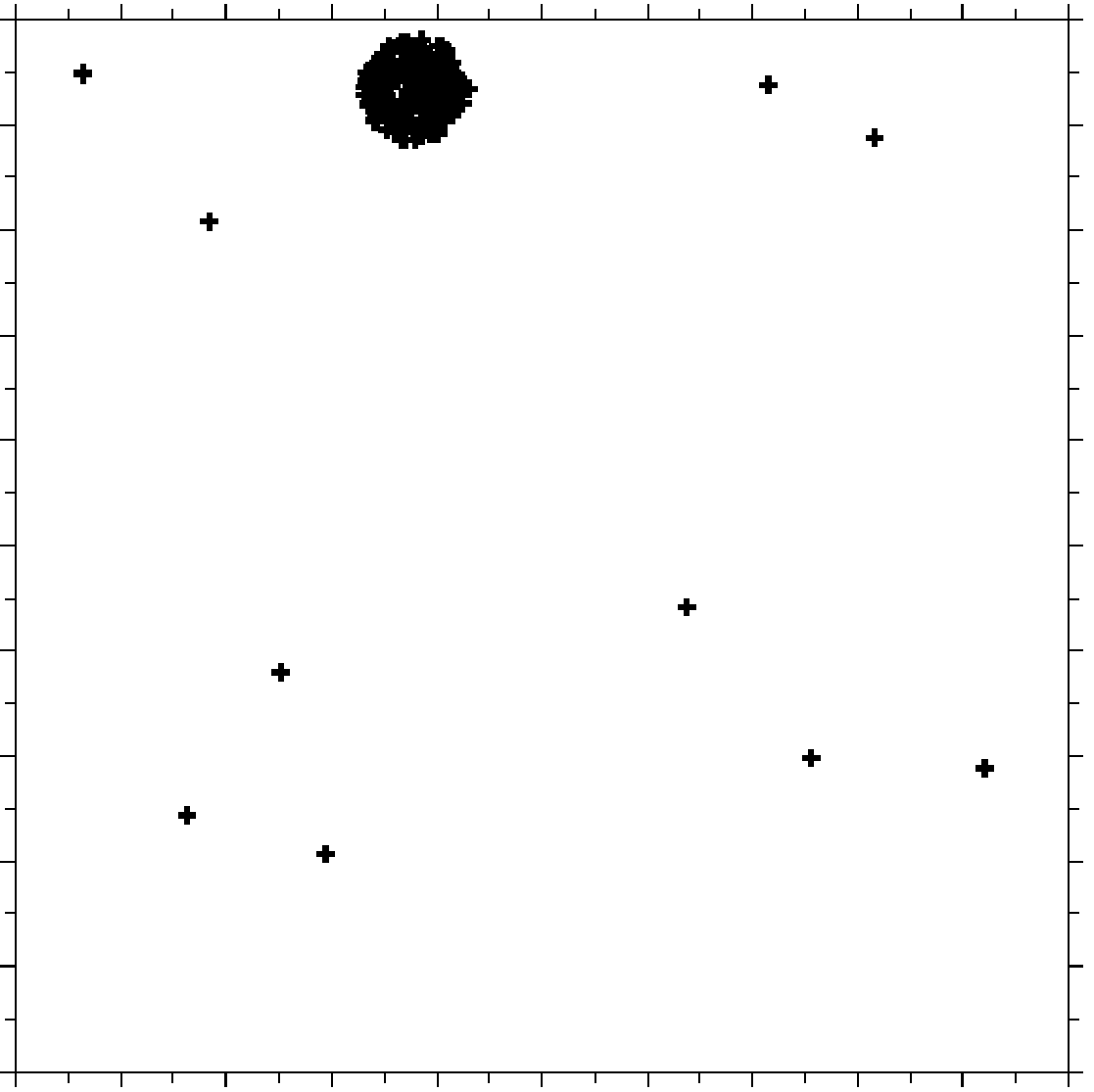}
\small{$t=1570.85 T_R$}
\end{minipage}
\caption{Formation of rotor clusters from 200 randomly distributed vortices with circulation
$\Gamma=2 \pi$ in a periodic box $L=40$. The fluctuations of the lattice of rotor clusters ends in a big
monopole. The time scale is given in the characteristic rotor turnover time $T_R$ of the system. }
\label{lattice} 
\end{figure*}

Our vortex model is based on the observation that the point vortex couple considered in section
\ref{models} under \emph{i.)} generates a far field that is similar to that of one elliptical vortex
with circulation $\Gamma$. We therefore consider point vortex couples with equal circulation
$\Gamma_i$ at the positions ${\bf x}_i$ and ${\bf y}_i$ as indicated in Fig. \ref{vector}. The
center of this object that we want to term a rotor is then given by ${\bf R}_i =\frac{{\bf x}_i
+{\bf y}_i}{2}$. In order to model a forcing and viscous damping mechanism similar to that
mentioned in section \ref{forcing}, the two point vortices in a rotor are supposed to be glued
together by an inelastic spring, such that each rotor possesses an additional degree of freedom and
that the size of a single rotor relaxes with relaxation time
$1/\gamma$ to $D_0$. Our model then reads   
\begin{eqnarray}\label{model}
\dot {\bf x}_i &=& 
\frac{\gamma}{2} (D_0-|{\bf x}_i-{\bf y}_i|) 
{\bf e}_i
+
\Gamma_i {\bf u}({\bf x}_i-{\bf y}_i)
\nonumber \\
&+& \sum_j \Gamma_j [
{\bf u}({\bf x}_i-{\bf x}_j)
+{\bf u}({\bf x}_i-{\bf y}_j)
]
\nonumber \\
\dot {\bf y}_i &=&- 
\frac{\gamma}{2} (D_0-|{\bf y}_i-{\bf x}_i|) 
{\bf e}_i
+
\Gamma_i {\bf u}({\bf y}_i-{\bf x}_i)
\nonumber \\
&+& \sum_j \Gamma_j [
{\bf u}({\bf y}_i-{\bf y}_j)+
{\bf u}({\bf y}_i-{\bf x}_j)]
\end{eqnarray}
where we have defined the unit vector
${\bf e}_i= \frac{{\bf x}_i-{\bf y}_i}{|{\bf x}_i-{\bf
    y}_i|}$ and the velocity field ${\bf u}({\bf r})$ is the velocity field
of a point vortex centered at the origin, ${\bf u}({\bf r})=
{\bf e}_z \times \frac{{\bf
    r}}{2\pi r^2}$.
The first two terms on the right-hand side of equation (\ref{model}) describe the interaction within
one rotor, whereas the last two terms describe the interaction with the other rotors. For vortices
moving inside a closed regime,
the velocity field has to be changed based on the introduction of
mirror vortices \cite{Saffman,Newton}.

It is important to stress that the above system is not a Hamiltonian system anymore due to the
inelastic coupling which mimics an energy input to the system on a scale $D_0$. Furthermore, by the
additional degree of freedom the rotor is sensitive with respect to a shear velocity field which can
be seen from the multipole expansion of the relative coordinate ${\bf r}_i= {\bf x}_i - {\bf y}_i$
with respect to the leading terms in $|{\bf r}|/|{\bf R}|$, derived in the appendix \ref{app}
\begin{eqnarray}\label{dipol1}\nonumber
\dot {\bf r}_i=\gamma(D_0 -r_i) \frac{{\bf r}_i}{r_i}
+2 \Gamma_i {\bf u}({\bf r}_i)
+
 \sum_{j}
\Gamma_j {\bf r}_i \cdot \nabla {\bf u}({\bf R}_{ij}) \\
\end{eqnarray} 
The influence of the forcing can be seen from the first term: If a rotor is subjected to shear, the 
spring between the point vortices in a rotor pulls back and the rotor relaxes to the size $D_0$. 
The shear velocity in the last term is thereby generated by the other rotors.\\ 
In a similar way, the multipole expansion of
the center coordinate of the rotor in appendix \ref{app} leads to the evolution equation 
\begin{equation}\label{locR}
\dot {\bf R}_i=2 \sum_j 
 \Gamma_j {\bf u}({\bf R}_{ij})
+\frac{1}{4} \sum_j
\Gamma_j  [({\bf r}_i \cdot \nabla)^2+({\bf r}_j \cdot \nabla)^2]
{\bf u}({\bf R}_{ij})  
\end{equation}

The evolution equation is identical to equation (\ref{x_i}), provided that the matrix $C_i$ can be
written as $C_i(t)= {\bf r}_i {\bf r}_i$, which corresponds to an infinitely thin elliptical vortex
oriented in ${\bf r}_i$-direction. The relative distance ${\bf r}_i$ can thus be considered as an
elliptical deformation of the velocity field that depends on the shear velocity field induced by the
remaining vortices and the effect of the overdamped spring. Furthermore, we again want to emphasize that the 
last term in Eq. (\ref{locR}) induces relative motions between the rotors as we have seen in 
section \ref{models}. The usual point vortex dynamics solely represented by the first term on the right hand 
side of equation (\ref{locR}) is thus extended to a dynamical system that is sensitive to the effect of
vortex thinning.

\section{Numerical results}

\begin{figure}[h]
\centering
\includegraphics[width=0.5 \textwidth]{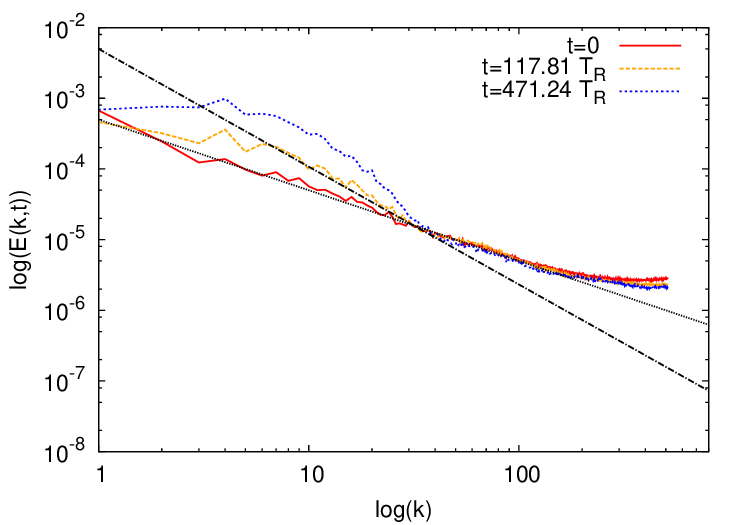}
\includegraphics[width=0.5 \textwidth]{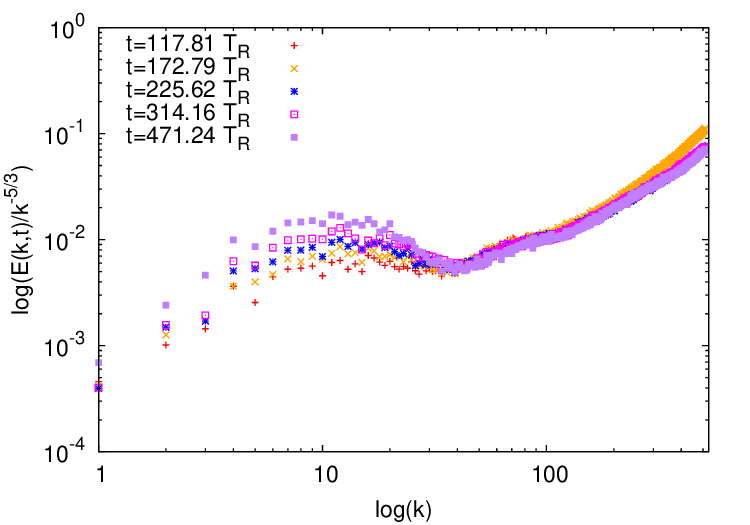}
\caption{(Color online) Above: Kinetic energy spectra of an assembly of rotors with $\Gamma+ \pm 2 \pi$ as in Fig. \ref{unequal} 
for different times $t$. Initially, the spectrum shows the charasteristic of a point vortex spectrum. 
The dotted line corresponds to $0.0005 k^{-1}$. As time evolves, the spectra begin to steepen corresponding to
an energy transfer into larger scales due to the effect of thinning. The dash-dotted line corresponds to $0.005 k^{-5/3}$.\\
Below: Compensated kinetic energy spectra from above. Only after $t\approx 100 T_R$, the spectra show the characteristic 
$5/3$-slope, which is in large part maintained over $\approx 300 T_R$.}
\label{spec1}
\end{figure}

We have numerically solved the dynamical system (\ref{model}) in a square periodic domain $L=40$. 
As a consequence of the periodic setting, the velocity kernels in (\ref{model}) have to be modified according to
\begin{equation}
 {\bf u}({\bf x}_i-{\bf x}_j)= {\bf e}_z \times \sum_{m,n=-N}^N \frac{{\bf x}_i-{\bf x}_j-mL{\bf e}_x-nL{\bf e}_y}
 {2 \pi|{\bf x}_i-{\bf x}_j-mL{\bf e}_x-nL{\bf e}_y|^2}
\end{equation}
where the boxes have been continued periodically, with up to $N=5$ layers of neighboring boxes, which
guarantees a sufficient degree of homogeneity. 
The temporal evolution of 200 rotors with an equal number of positive and negative circulations starting
from a random initial condition exhibits the formation of a large scale vortical 
structure via the formation of rotor-clusters. 

A typical time series is exhibited in Fig. \ref{unequal}, for the parameter values 
($N=200$, $\Gamma=\pm 2 \pi$, $\gamma=5.$, $D_0=.5$, $L=40$). 
The temporal evolution of the system can be quantified 
by the introduction of a characteristic time scale of the system which is given as the period that a rotor 
possesses at a fixed distance $D_0$ and is in the following termed as one rotor turnover time 
$T_R=\frac{\Gamma}{2\pi^2 D_0^2}$, which follows from equation (\ref{Gauss}) for the case of vanishing viscosity.

\begin{figure}[h]
\centering
  \setbox1=\hbox{\includegraphics[width=0.5 \textwidth]{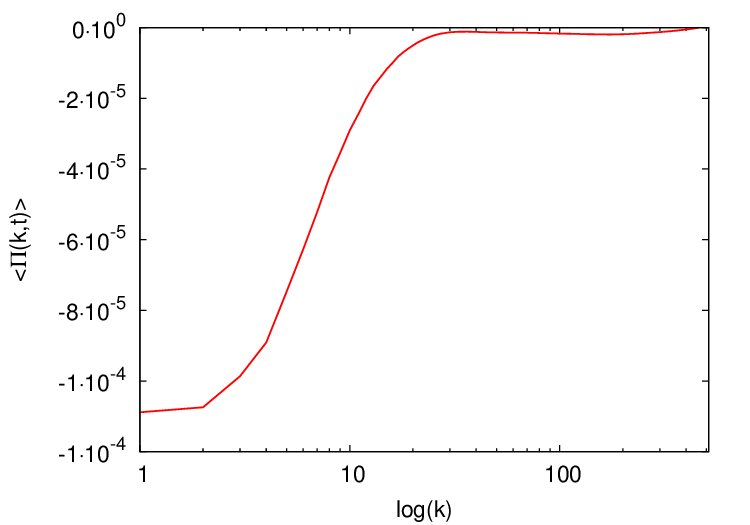}}
  \includegraphics[width=0.5 \textwidth]{pi.ps}\llap{\raisebox{1cm} {\includegraphics[width=0.25 \textwidth]{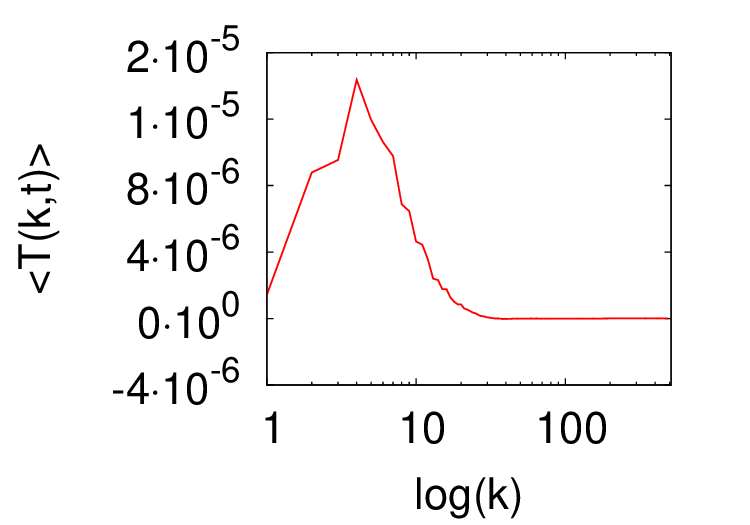}}\hspace{0.6cm}}
\caption{(Color online) Time-averaged kinetic energy flux $\langle \Pi(k,t) \rangle$ calculated from the spectra with $5/3$-slope in Fig. \ref{spec1}. 
An energy transfer into the large scales can clearly be observed. The inlet corresponds to the time-averaged kinetic 
energy transfer rate $\langle T(k,t) \rangle$.  }
\label{flux}
\end{figure}

\begin{figure}[h]
\centering
\includegraphics[width=0.5 \textwidth]{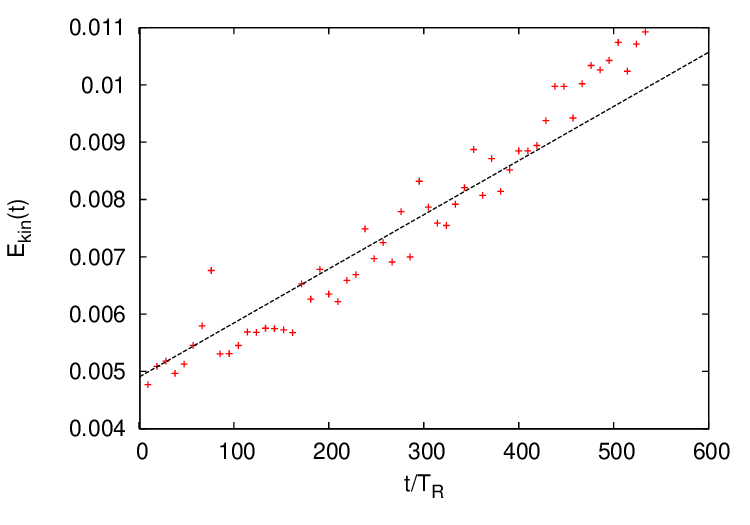}
\caption{(Color online) Temporal evolution of the kinetic energy of the binary rotor system from Fig. \ref{unequal}. The fitted line possesses 
a slope $\epsilon=(0.908 \pm 0.052)\cdot 10^{-4}$.}
\label{energy}
\end{figure}

As it can be seen from Fig. \ref{unequal}, the clustering of like-signed rotors already occurs within the first 
100 rotor turnover times, which means that the separation of the rotors takes place on a relatively short time-scale.
The temporal evolution of 200 rotors with identical circulations starting from random initial positions of the
rotors, is exhibited in Fig. \ref{lattice}. A fluctuating lattice of rotor clusters appears and after approximately 
1500 rotor turnover times, the system forms a monopole which attracts the remaining rotors.

We have calculated the kinetic energy spectra of the rotor system with $\Gamma = \pm 2 \pi$ at different times in  
Fig. \ref{spec1}. Starting from 20 different initial configurations of the rotors, we let the 
systems evolve in time and performed the ensemble average at a specific time $t$. Thereby, the spectrum is calculated
from the velocity field in Eq. (\ref{biot}) that has been interpolated on a grid and then transformed into Fourier space.

Initially, the rotors possess a clear point vortex spectrum following a power law $E(k,t=0) \sim k^{-1}$. Only at high
values of $k$ deviations due to the singular structure of the vorticity and corresponding discontinuities in the velocity 
field manifest themselves in an increase of $E(k,t)$. This effect can be observed in the following spectra, too.
However, after a few ($\approx 100$) rotor turnover times, as the rotor clustering sets in, a more universal energy spectrum
can be observed. Due to an energy flux from smaller to larger scales, the spectra begin to steepen for smaller $k$-values, 
revealing a spectrum that is close to the predicted $E(k,t) \sim k^{-5/3}$. As it can be seen from the compensated spectra in
Fig. \ref{spec1}, this slope remains constant for nearly $500 T_R$ and an energy flux into the large scales takes place.
This is also in agreement with the time-averaged spectral energy flux $\Pi(k,t)$, depicted in Fig. \ref{flux}. The inlet plot in Fig. \ref{flux}
corresponds to the kinetic energy transfer rate $T(k,t)$, which is related to $\Pi(k,t)$ according to \cite{Vincent}
\begin{equation}
 \Pi(k,t)= \int_k^{\infty} \textrm{d}k' T(k',t)
\end{equation}
It is obvious
that energy accumulates at small $k$-values. This is not surprising, since the rotor model only provides an energy input 
on small scales and it will be a task for the future to extend the model in order to achieve a damping at small values of $k$
and thus to extract energy at the integral scale.

We now turn to the determination of the Kolmogorov constant of the energy spectrum from the binary rotor system 
($\Gamma = \pm 2 \pi$). The spectrum as it was predicted by Kraichnan \cite{Kraichnan1} reads
\begin{equation}
 E(k)=C_K \epsilon^{2/3} k^{-5/3}
\end{equation}
where $\epsilon$ is the energy dissipation rate.

In the following, $\epsilon$ is determined from the time-dependence
of the total kinetic energy that shows up to be linear in time within $\approx 5 $\%, proving that the model is not
Hamiltonian anymore due to the inclusion of the forcing term. The corresponding plot is 
depicted in Fig. \ref{energy}.  

The slope of the fitted line can thus be interpreted as the rate of energy input into the system and we obtain a value of
$\epsilon=(0.908 \pm 0.052)\cdot 10^{-4}$. In order to make an estimate for $C_K \epsilon^{2/3}$, we take an average of the 
compensated spectra in Fig. \ref{spec1} of times between $120 T_R$ and $620T_R$ which yields $C_K \epsilon^{2/3}=0.0119 \pm 0.0015$.
The Kolmogorov constant $C_K$ of the rotor system for times t between $120 T_R$ and $620 T_R$ thus lies in the range
$C_K= 5.89 \pm 0.74$. The high inaccuracy of our estimate is due to the estimation of  $C_K \epsilon^{2/3}$.  
Reported values from direct numerical simulations \cite{Uriel,Yakhot,Bofetta} and experiments \cite{Tabeling1,Tabeling2} 
lie within the range from 5.8 to 7.0. 
The Kolmogorov constant of the rotor system thus lies on the lower end of that range. 
In comparison to the point vortex model of Siggia and Aref \cite{Siggia}, who report a 
Kolmogorov constant of $C_K=14$ which is twice the accepted value, the rotor model thus seems to provide an efficient mechanism
for the energy transfer upscale due to the effect of vortex thinning. 

Another important way to determine the distribution and the occuring structures in the rotor model 
will be discussed in the following.
In order to quantify the emergence of the rotor clusters in Fig. \ref{unequal} and \ref{lattice}, we make use of the radial
distribution function $g({\bf r})$ which can be considered as the probability of finding
a like-signed rotor at a distance ${\bf r}$ away from a reference-rotor (for further references see for instance \cite{toda}). 
The radial distribution function is therefore given as  
\begin{equation}
 g({\bf r})= \frac{1}{\rho} \left \langle  \sideset{}{'} \sum_{i,j} \delta({\bf x}_i-{\bf x}_j-{\bf r}) \right \rangle  
\end{equation}
where $\rho=\frac{L^2}{N^2}$ and the prime indicates that summation over $i =j$ is left out. The averaging is performed in such 
a way that the number of like-signed rotors populating a concentric segment of radius $\textrm{d}r$ at a given radius r is 
divided by its area. In the following the radial distribution function is assumed to be isotropic, so that $g({\bf r})=g(r)$.
For a disordered state one expects the radial distribution function to be equal to 1 for every
${r}$. As the formation of the rotor clusters sets in, one should observe an increase of $g(r)$
for small $r$, since the probability of finding a like-signed rotor in the neighborhood of a
reference-rotor increases.
\begin{figure}
\centering
\includegraphics[width=0.5 \textwidth]{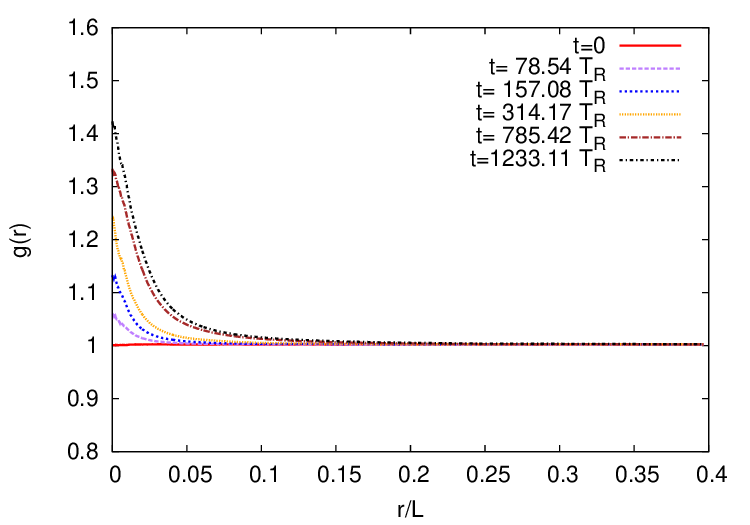}
\includegraphics[width=0.5 \textwidth]{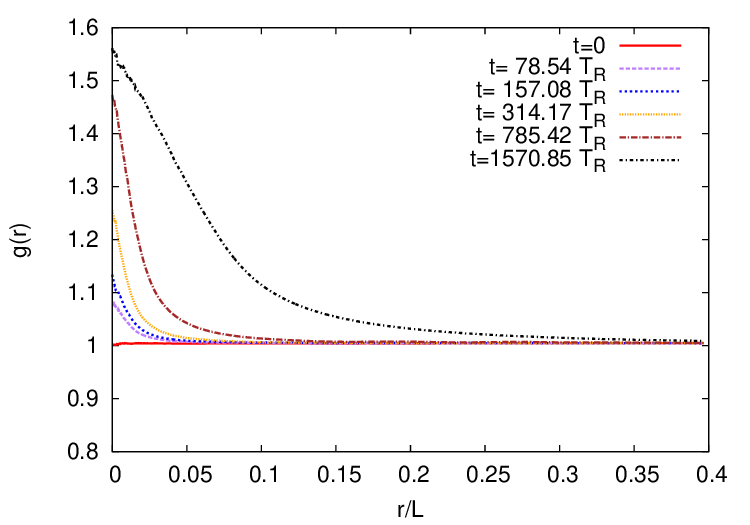}
\caption{(Color online) Above: Radial distribution functions for six different times from the time series of rotors with 
$\Gamma= \pm 2 \pi$ in Fig. \ref{unequal}. An increase of $g(r)$ for small $r$ can be observed meaning an 
increased probability of finding a like-signed rotor in the neighborhood of a reference-rotor.\\
Below: Radial distribution functions for six different times from the time series of the rotors with $\Gamma=2 \pi$ 
in Fig. \ref{lattice}. Again, an increase of $g(r)$ can be observed. 
The formation of the final monopole manifests itself in a long-ranging $g(r)$.}
\label{radial_dis}  
\end{figure}

\begin{figure}[h!]
\centering
\includegraphics[width=0.5 \textwidth]{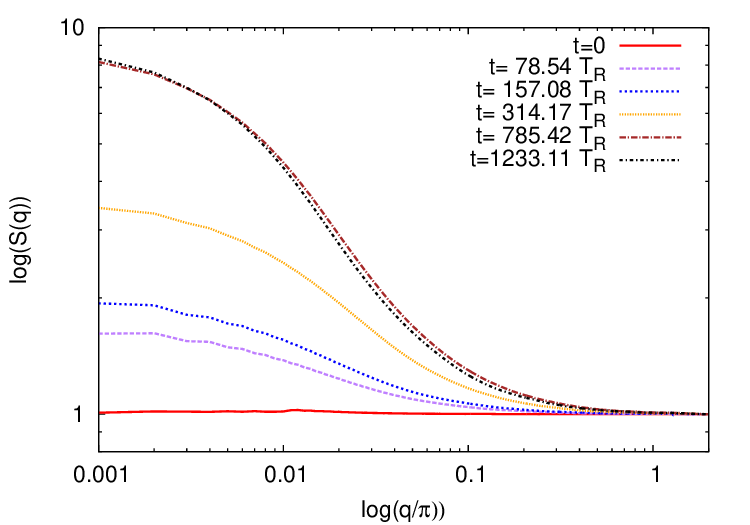}
\includegraphics[width=0.5 \textwidth]{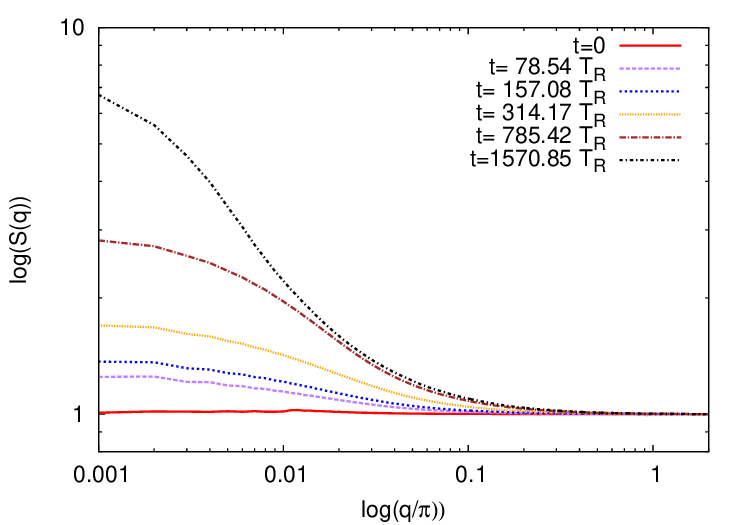}
\caption{(Color online) Above: Structure factors $S(q)$ from the radial distribution function of the time series of rotors with $\Gamma= \pm 2 \pi$ in Fig. \ref{unequal}.\\
 Below: Structure factors $S(q)$ from the radial distribution function of the time series of rotors with $\Gamma= 2 \pi$ in Fig. \ref{lattice}. The flucuations of the rotor lattice is accompagnied by an increase of the structure factor $S(q)$ over time.}
\label{structure}
\end{figure}
\begin{figure}[h]
\centering
\includegraphics[width=0.5 \textwidth]{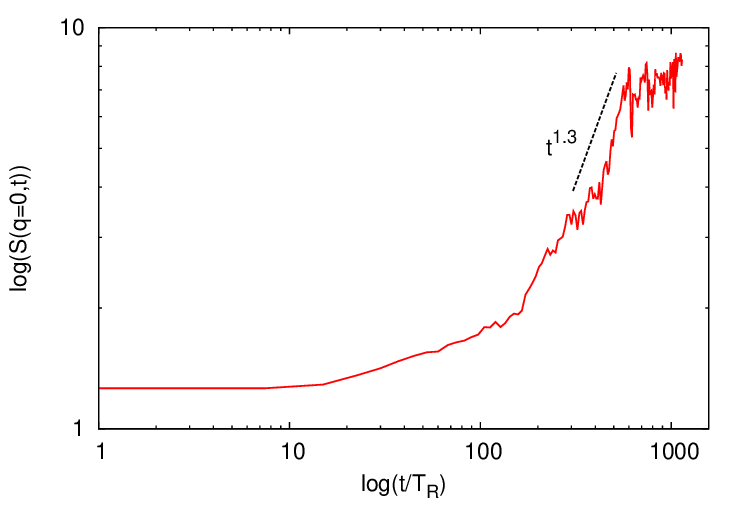}
\includegraphics[width=0.5 \textwidth]{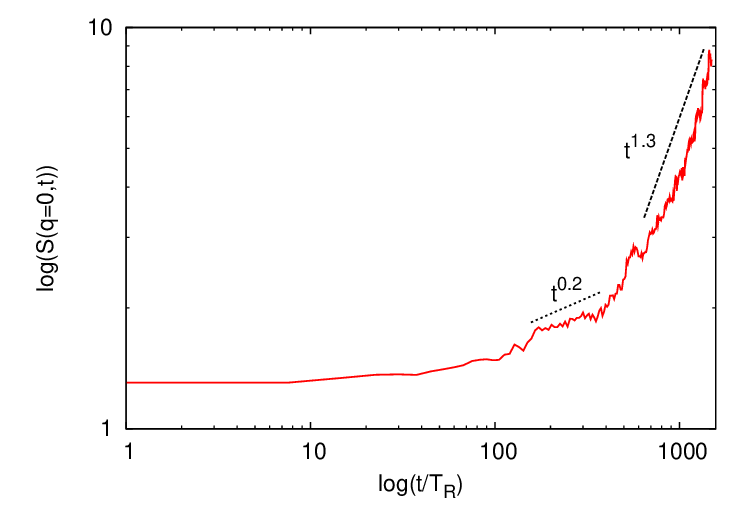}
\caption{(Color online) Above: Temporal evolution of $S(q=0,t)$ of the time series of rotors with $\Gamma= \pm 2 \pi$ in Fig. \ref{unequal}. A power law with $t^{1.3}$ is plotted for comparison. The strong growth rate saturates after $t \approx 500 T_R$.\\
 Below: Temporal evolution of $S(q=0,t)$ of the time series of rotors with $\Gamma= 2 \pi$ in Fig. \ref{lattice}. Two power laws $t^{0.2}$ and $t^{1.3}$ are plotted for comparison.}
 \label{temp}
\end{figure}

The radial distribution functions for the two time series are plotted in
Fig. \ref{radial_dis}  and one clearly observes an increase of $g(r)$ at small $r$. In order to get smooth
curves, $g(r)$ was calculated in such a way that it shows no
discontinuities for $r \approx 0$ due to a minimum distance between neighboring rotors. The radial
distribution function can thus be used as a qualitative measure for the formation of the clusters
and their typical sizes.  
Furthermore, the radial distribution function is related to the structure factor 
\begin{equation}
 S({\bf q})= 1 + \frac{1}{N} \left \langle \sideset{}{'}  \sum_{i,j} e^{-i {\bf q} \cdot ({\bf x}_i -{\bf x}_j)}
\right \rangle
\end{equation}
in a way that 
\begin{eqnarray}\label{structurefactor} \nonumber
 S({\bf q})&=&1+ \rho \int \textrm{d} {\bf r}  (g(r)-1)e^{-i {\bf q} \cdot {\bf r}}\\ 
 &=&1+ 2 \pi \rho \int_{0}^{\infty} \textrm{d} r
~r (g(r)-1) J_0(qr)
\end{eqnarray}
where $J_0(qr)$ is the Bessel function of order zero. The structure factor $S(q)$  can thus be calculated via the Hankel 
transform of $(g(r)-1)$, provided that the radial distribution function is isotropic. The structure factors for the two system 
are plotted in Fig. \ref{structure}. For the case of the mixed system of Fig. \ref{unequal}, one observes an increase 
of $S(q)$ over time. Whether this increase is governed by a power law for intermediate $q$ has to be evaluated within 
further simulations of the model equations (\ref{model}). Furthermore, Eq. (\ref{structurefactor}) is of great importance
for the investigation of the rotor model, since it relates macroscopic quantities on the left-hand side to microscopic quantites 
such as the radial distribution function. It is thus a good starting point for the interpretation of the fluctuations of the rotor 
clusters in the realm of phase transitions.      

The growth rate of the rotor clusters can be determined from the time dependence of the structure factor. 
The growth of the largest structures of the system is given by $S(q=0,t)$. In Fig. \ref{temp}, the temporal evolution 
of $S(q=0,t)$ is plotted for the two systems.  
The fluctuating rotor lattice below exhibits a pronounced growth rate after $t \approx 600 T_R$, whereas the 
growth rate of the mixed system above already increases for $t \approx 150 T_R$.  For comparison, two power 
laws $t^{0.2}$ and $t^{1.3}$ were plotted in the figures. The growth rate of our rotor clusters can thus be considered as 
relatively strong compared to typical growth rates from pattern formation, for instance compared to 
the growth rate of droplets in the Cahn-Hilliard equation where $S(q=0,t) \sim t^{1/3}$ according to 
Slyozov-Lifshitz theory \cite{Lifshitz}.   
   
The fact that the rotor vortex system exhibits a pronounced 
inverse cascade already for 
moderate numbers of rotors (200 rotors have been used for the figures) on a small time-scale
allows us to investigate the inverse cascade
using methods of nonlinear dynamics.
Although, usual point vortex models such as \cite{Siggia}, have been known for a long time to possess 
inverse energy cascades the present model incorporates the aspect of vortex thinning, due
to a possible change of the ellipticity of the rotor in much the same way
as identified in the experiments of Chen et al \cite{Ecke}.  
Hence, it is a minimal
dynamical model containing the mechanisms of the inverse cascade.
In the following we shall discuss the origin of the formation of 
clusters of rotors with like-signed circulations. 

\section{ Interaction of two rotors}\label{inter}
As it has been discussed in section \ref{models}, the deformation 
of the shapes of the vortices induces relative motion between their centers. However, the analytical 
calculation of these relative velocities directly from the fluid dynamical equations as it has been
performed for instance in \cite{Melander}, are quite difficult to handle in order to get meaningful
results of the dynamics underlying the inverse cascade. However, since our model possesses two kinds of 
dynamics, i.e. a fast dynamic within the rotation of the point vortex pairs in one rotor and a slower
dynamic within the interaction between the rotors, it is possible to simplify the corresponding equations
within an adiabatic approximation of the fast rotations. The result for the relative motion between 
two rotors reveals the importance for the dynamical aspects of the inverse cascade caused by an attractive
motion in between two like-signed vortices and the 
symmetry-breaking $\Gamma_i \rightarrow -\Gamma_i$ due to the introduction of the forcing in 
Eq. (\ref{model}). 

In the following, we consider the configuration of two rotors with circulations $\Gamma_i$ and
$\Gamma_j$, depicted in Fig. \ref{vector}, which can be considered as the interaction of two
infinitely-thin elliptical vortices in the same manner as \emph{ii.)} from section \ref{models}.  
It is straightforward to show that the center of vorticity
$
\frac{(\Gamma_i {\bf R}_i+
\Gamma_j {\bf R}_j)}{\Gamma_i+\Gamma_j}
$
is a conserved quantity. The distance vector ${\bf R}_{ij}={\bf R}$ between the two rotors
obeys the evolution equation   
\begin{eqnarray}
\dot {\bf R} &=&  \frac{(\Gamma_i+\Gamma_j)}{\pi}
{\bf e}_z \times  \left[\frac{{\bf R}}{|{\bf R}|^2}
\right. \\ \nonumber
&~&+
\frac{1}{8} [
 -2 \frac{\bf R}{|{\bf R}|^4} {\bf r}_i^2-
4 \frac{{\bf r}_i}{|{\bf R}|^4}{\bf r}_i\cdot {\bf R}+8 
\frac{{\bf R}}{|{\bf R}|^6} ({\bf r}_i\cdot {\bf R})^2 ]
 \\ \nonumber
&~&+  
\frac{1}{8}\left.[
 -2 \frac{\bf R}{|{\bf R}|^4} {\bf r}_j^2-
4 \frac{{\bf r}_j}{|{\bf R}|^4}{\bf r}_j\cdot {\bf R}+8 
\frac{{\bf R}}{|{\bf R}|^6} ({\bf r}_j\cdot {\bf R})^2 ]
\right]
\end{eqnarray}
which follows from equation (\ref{locR}) in calculating the corresponding velocity field gradients,
described in the appendix \ref{app}.\\
For the following it is convenient to represent the unit vectors according to ${\bf e}_i=\frac{{\bf
r}_i}{r_i}=\left( \begin{array}{l}     
    \cos \varphi_i  \\
    \sin \varphi_i \end{array} \right)$, as well as
 ${\bf e}_R=\frac{{\bf R}}{R}=\left( \begin{array}{l}     
    \cos \varphi_R  \\
    \sin \varphi_R \end{array} \right)$, which yields 
\begin{equation}
{\bf e}_R \cdot 
[{\bf e}_z \times {\bf e}_i]({\bf e}_i\cdot {\bf e}_R)=\frac{1}{2} \sin(2 (\varphi_i -\varphi_R)) 
\end{equation} 
 We obtain the equation for the relative
distance
\begin{eqnarray}\label{R_i}\nonumber
\dot R = -\frac{(\Gamma_i+\Gamma_j)}{4\pi} \frac{1}{R^3}
&[&r_i^2 \sin(2 (\varphi_i -\varphi_R)) \\
&+&r_j^2 \sin(2 (\varphi_j -\varphi_R))
]
\end{eqnarray}

The evolution equation for the relative coordinate of a rotor
reads
\begin{eqnarray}\nonumber
\dot {\bf r}_i &=& 
\gamma(D_0-r_i) \frac{{\bf r}_i}{r_i}
+
 \frac{\Gamma_i}{\pi} {\bf e}_z \times \frac{{\bf r}_i}{r_i^2}\\ 
&~&+\frac{\Gamma_j}{2\pi}
{\bf e}_z \times \left[\frac{{\bf r}_i}{|{\bf R}|^2}-2 \frac{{\bf R}}{|{\bf R}|^4}
({\bf r}_i\cdot {\bf R}) \right]
\end{eqnarray}
which follows from equation (\ref{dipol1}) and the calculation of the velocity field gradients,
performed in the appendix \ref{app}.
We have to determine the quantities 
$r_i^2$, $r_j^2$, which are determined by the evolution equations
\begin{eqnarray}
\dot r_i=\gamma(D_0-r_i)+ \frac{\Gamma_j}{2\pi} \frac{r_i}{R^2} \sin(2 (\varphi_i-\varphi_R))
\nonumber \\
\dot r_j=\gamma(D_0-r_j)+ \frac{\Gamma_i}{2\pi} \frac{r_j}{R^2} \sin(2 (\varphi_j-\varphi_R))
\end{eqnarray} 
We can solve iteratively for small deviations of $r_i$ from $D_0$:
\begin{equation}
r_i=D_0+ \frac{\Gamma_j}{2\pi} \frac{D_0}{R^2}
\int_{-\infty}^t \textrm{d}t' e^{-\gamma(t-t')} \sin(2 (\varphi_i(t')-\varphi_R(t')))
\end{equation}
A similar treatment applies to $r_j$. 
Splitting the rotation into its fast ($\sim e^{2i (\omega_i -\omega_R)t}$) and slow 
varying parts $e^{2i(\tilde \varphi_i(t)-\tilde \varphi_R(t))}$, i.e. 
\begin{equation}
 e^{2i(\varphi_i(t)-\varphi_R(t))}\approx e^{2i (\omega_i -\omega_R)t} 
 e^{2i(\tilde \varphi_i(t)-\tilde \varphi_R(t))}
\end{equation}
 we obtain after a partial integration
\begin{eqnarray}\label{adiabatic} \nonumber
&~& \int_{-\infty}^t \textrm{d}t' e^{-\gamma(t-t')} e^{2i (\tilde \varphi_i(t')-\tilde \varphi_R(t'))}\\ \nonumber 
&~&= \frac{ e^{2i (\omega_i -\omega_R)t} e^{2i(\tilde \varphi_i(t)-\tilde \varphi_R(t))}}
{2i (\omega_i-\omega_R)+ \gamma}\\ \nonumber 
&~&- \int_{-\infty}^t \textrm{d}t' e^{-\gamma(t-t')+2i (\omega_i -\omega_R)t'}  \left[ (\omega_i-\omega_R)+ \frac{\gamma}{2i}\right]^{-1}\\ \nonumber
&~&\times e^{2i(\tilde \varphi_i(t')-\tilde \varphi_R(t'))}(\dot{\tilde {\varphi}}_i(t')-\dot {\tilde {\varphi}}_R(t'))\\
~
\end{eqnarray}
In order to proceed with the adiabatic approximation, we neglect the second term in Eq. (\ref{adiabatic}) since it contains time derivatives of the slowly varying parts of the rotations. Assuming that the damping constant
$\gamma$ is large compared to the rotation frequency of the rotor, we obtain 
\begin{eqnarray}\nonumber
&~&\int_{-\infty}^t \textrm{d}t' e^{-\gamma(t-t')} \sin(2 (\varphi_i(t')-\varphi_R(t')))\\  
&~& \approx
\frac{\sin(2 (\varphi_i(t)-\varphi_R(t)))}{\gamma}
\end{eqnarray}
To lowest order in $\gamma^{-1}$ we thus obtain
\begin{eqnarray}\nonumber
r_i^2 &=& D_0^2 \left[1+ \frac{\Gamma_j}{\pi \gamma R^2}
\sin(2 (\varphi_i-\varphi_R)) \right]
\nonumber \\
r_j^2 &=& D_0^2\left[1+ \frac{\Gamma_i}{\pi \gamma R^2}
\sin(2 (\varphi_j-\varphi_R))\right]
\end{eqnarray}
Here, the last terms on the right-hand side arise due to the 
change of the size of the rotors, connected with a change of the far field,
induced by the mutually generated shear. It thus mimics the mechanism of vortex
thinning, identified in \cite{Ecke}.\\
The relative motion of the rotors
obeys the evolution equation
\begin{eqnarray} \nonumber
\dot R &=& -\frac{(\Gamma_i+\Gamma_j)}{4\pi} \frac{D_0^2}{R^3}
 \bigg[\sin(2 (\varphi_i-\varphi_R))+\sin(2 (\varphi_j-\varphi_R)) \\ \nonumber
&~& +  \frac{\Gamma_j}{\pi \gamma R^2} \sin^2(2 (\varphi_i-\varphi_R))
+ \frac{\Gamma_i}{\pi \gamma R^2} \sin^2(2 (\varphi_j-\varphi_R))\bigg] 
\end{eqnarray}
We now average the evolution equation with respect to the rotations
of the vectors ${\bf e}_i(t)$ and ${\bf e}_j(t)$ taking into account that the averages
$\langle \sin(2 (\varphi_i-\varphi_R))\rangle=\langle\sin(2 (\varphi_j-\varphi_R)) \rangle=0$
vanish.
Furthermore, the averages
$\langle \sin^2(2 (\varphi_i-\varphi_R)) \rangle=\langle\sin^2(2 (\varphi_j-\varphi_R))\rangle =a $
are positive. 
As a consequence, the relative distance behaves according to
\begin{equation}\label{R}
\dot R=- \frac{(\Gamma_i+\Gamma_j)^2}{(2\pi)^2} \frac{D_0^2}{\gamma R^5}  a
\end{equation}
Two rotors approach each other, except for $\Gamma_i =-\Gamma_j$. 
It is important to stress that this
attractive relative motion arises only if we include the irreversible effect
of the strain induced stretching of the rotors. Furthermore, the symmetry breaking of $\Gamma_i
\rightarrow -\Gamma_i$ in equation (\ref{R}) can be considered as an important feature of the rotor
model in comparison to the point vortex model, which conserves this symmetry.

\section{Conclusions} We have presented a generalized 
point vortex model, a rotor model,
exhibiting an inverse 
cascade based on clustering of rotors. We have discussed how this rotor model
can be derived from the vorticity equation by an expansion of the vorticity
field into a set of elliptical vortices at locations ${\bf x}_i(t)$ and 
shapes $C_i(t)$. An important point has been the inclusion of a forcing term,
which prevents the elliptical far field of the rotors
from diffusing away.  
The added forcing term breaks the symmetry 
$\Gamma_i \rightarrow -\Gamma_i$, $t \rightarrow -t$. This
symmetry breaking lies at the origin of cluster formation and the
inverse cascade, as can be 
seen from the two-rotor interaction inducing in average a relative motion
proportional to $\frac{D_0^2}{\gamma} (\Gamma_1+\Gamma_2)^2/R^5$. 

The numerical simulations of the model equations (\ref{model}) reveal the formation
of rotor clusters on a short time scale. In addition, the calculated energy spectra and 
energy fluxes give strong evidence for the important role of vortex thinning during the 
cascade process in two-dimensional turbulence. 

The presented rotor model can be investigated by applying methods from dynamical
systems theory like the evaluation of finite time Ljapunov exponents and
Ljapunov vectors. These and further dynamical aspects are the basis for future work and will be
covered in a following paper. The model system (\ref{modelg1}) may also be
studied as a stochastic system by considering the velocity ${\bf U}_i(t)$ to
be a white noise force. The corresponding Fokker-Planck equation allows one
to draw analogies with quantum mechanical many body problems.  Furthermore, we
emphasize that a continuum version of the model equations (\ref{modelg1}) leads
to a subgrid model exhibiting analogies with
the work of Eyink \cite{Eyink}.  

It will be a task
for the future to investigate the cluster formation from a statistical point
of view, based on the formulation of kinetic equations, along the lines as has
been performed for fully developed turbulence \cite{Wilczek1,Wilczek2,Wilczek3}, and Rayleigh-B\'enard 
convection \cite{Lulff}. 
In this respect we
hope to find a relation to the kinetic equation for the two-point vorticity 
statistics recently derived on the basis of
the Monin-Lundgren-Novikov hierarchy, taking conditional averages from direct
numerical simulations \cite{arXiv}.  
\begin{acknowledgments}
J.F. is very grateful for discussions with Michael Wilczek and Frank Jenko about the organization of
this paper. Sadly, Rudolf Friedrich (\dag 16th August 2012) unexpectedly passed away during this
work. He was as much an inspiring physicist as well as a caring father.  
\end{acknowledgments}

\appendix
\section{}\label{app}
In this part we calculate the multipole expansion of a rotor, defined by ${\bf x}_i$ and ${\bf y}_i$
in Fig. \ref{vector}. To this end, we introduce relative and center coordinates, according to 
\begin{equation}
 {\bf r}_i = {\bf x}_i -{\bf y}_i \qquad \textrm{and} \qquad {\bf R}_i =\frac{{\bf x}_i +{\bf y}_i}{2}
\end{equation}
as well as the vector ${\bf R}_{ij}= {\bf R}_i -{\bf R}_j$.\\
In using equation (\ref{model}), we obtain the evolution equation for the relative coordinate
\begin{eqnarray} \nonumber
 \dot {\bf r}_i  &=& \gamma (D_0 -r_i){\bf e}_i + 2 \Gamma_i {\bf u}({\bf r}_i)\\ \nonumber
 &~&+ \sum_j \Gamma_j \left \{ {\bf u} \left({\bf R}_{ij} + \frac{{\bf r}_i-{\bf r}_j}{2} \right)  + {\bf u} \left({\bf R}_{ij} + \frac{{\bf r}_i+{\bf r}_j}{2} \right) \right.\\
&~&\left.- {\bf u} \left({\bf R}_{ij} - \frac{{\bf r}_i+{\bf r}_j}{2} \right) - {\bf u} \left({\bf R}_{ij} - \frac{{\bf r}_i-{\bf r}_j}{2} \right) \right \}
\end{eqnarray}
A Taylor expansion of the curled bracket yields
\begin{eqnarray}
 \dot {\bf r}_i &=& \gamma (D_0 -r_i){\bf e}_i + 2 \Gamma_i {\bf u}({\bf r}_i) \\ \nonumber
 &~& +\sum_j
\Gamma_j\left \{ ({\bf r}_i -{\bf r}_j) \cdot \nabla_{{\bf R}_{ij}} + ({\bf
r}_i +{\bf r}_j) \cdot \nabla_{{\bf R}_{ij}} \right \} {\bf u}({\bf R}_{ij}) \\ \nonumber
&=& \gamma (D_0 -r_i){\bf e}_i + 2 \Gamma_i {\bf u}({\bf r}_i)+ 2 \sum_j \Gamma_j {\bf r}_i \cdot \nabla_{{\bf R}_{ij}} {\bf u}({\bf R}_{ij} )    
\end{eqnarray}
where we have only retained the leading terms in $|{\bf r}|/|{\bf R}|$.  
The evolution equation for the center coordinate reads
\begin{eqnarray} \nonumber
 \dot {\bf R}_i  &=&\frac{1}{2}  \sum_j \Gamma_j \left \{ {\bf u} \left({\bf R}_{ij} + \frac{{\bf r}_i-{\bf r}_j}{2} \right) + {\bf u} \left({\bf R}_{ij} + \frac{{\bf r}_i+{\bf r}_j}{2} \right) \right.\\ 
&~& \left. + {\bf u} \left({\bf R}_{ij} - \frac{{\bf r}_i+{\bf r}_j}{2} \right) + {\bf u} \left({\bf R}_{ij} - \frac{{\bf r}_i-{\bf r}_j}{2} \right) \right \}
\end{eqnarray}
Again, a Taylor expansion yields
\begin{eqnarray}\nonumber
 &\dot {\bf R}_i&=2 \sum_j \Gamma_j \bigg [ {\bf u}( {\bf R}_{ij}) \\ \nonumber
 &+& \frac{1}{8} \left \{ [
({\bf r}_i - {\bf r}_j ) \cdot \nabla_{{\bf R}_{ij}} ]^2 + [ ({\bf r}_i + {\bf r}_j ) \cdot
\nabla_{{\bf R}_{ij}} ]^2 \right \}  {\bf u}( {\bf R}_{ij}) \bigg] \\ \nonumber
&=& 2 \sum_j \Gamma_j  {\bf u}( {\bf R}_{ij})\\ 
&~&+ \frac{1}{4} \sum_j \Gamma_j [ ({\bf r}_i \cdot \nabla_{{\bf R}_{ij}})^2 +({\bf r}_j \cdot \nabla_{{\bf R}_{ij}})^2 ] {\bf u} ({\bf R}_{ij})
\end{eqnarray}
 The gradients of the velocity fields are now calculated according to
\begin{equation}
{\bf r}\cdot \nabla_{\bf R} \frac{{\bf R}}{|{\bf R}|^2}
= \frac{{\bf r}}{|{\bf R}|^2}-2  \frac{{\bf R}}{|{\bf R}|^4}
{\bf r}\cdot {\bf R}
\end{equation}
which is needed in equation (\ref{dipol1}), and
\begin{equation}
({\bf r}\cdot \nabla_{\bf R})^2 \frac{{\bf R}}{|{\bf R}|^2}
= -2 \frac{\bf R}{|{\bf R}|^4} {\bf r}^2-
4 \frac{{\bf r}}{|{\bf R}|^4}{\bf r}\cdot {\bf R}+8 
\frac{{\bf R}}{|{\bf R}|^6} ({\bf r}\cdot {\bf R})^2
\end{equation}
Now, this is the counterpart of equation (\ref{locR}).

\end{document}